\documentclass[12pt]{article}
\usepackage{mathpazo}
\usepackage[T1]{fontenc}
\usepackage[latin9]{inputenc}
\usepackage{geometry}
\usepackage{babel}
\usepackage{float}
\usepackage{url}
\usepackage{amsmath}
\usepackage{amsthm}
\usepackage{amssymb}
\usepackage{undertilde}
\usepackage{xspace}
\usepackage{microtype}
\usepackage{enumitem}
\usepackage{booktabs}

\usepackage[unicode=true,pdfusetitle, bookmarks=true,bookmarksnumbered=false,bookmarksopen=false,breaklinks=true,pdfborder={0 0 1},backref=false,colorlinks=true, citecolor=blue] {hyperref}

\makeatletter

\usepackage{bm}
\usepackage{extarrows}
\usepackage{mathtools}
\usepackage{bbm}
\usepackage{commath}
\usepackage{cases}
\usepackage{algorithm}
\usepackage{subcaption}
\usepackage{algorithmic}

\newtheorem{theorem}{Theorem}

\usepackage{appendix,breakurl}
\PassOptionsToPackage{hyphens}{url}

\makeatother

\begin{document}

\title{Universal Policy Tracking: Scheduling for Wireless Networks with Delayed State Observation}

\author{Bai Liu, Eytan Modiano \\
\normalsize Massachusetts Institute of Technology
}

\date{}                     

\maketitle

\begin{abstract}
  Numerous scheduling algorithms have been proposed to optimize various performance metrics like throughput, delay and utility in wireless networks. However, these algorithms often require instantaneous access to network state information, which is not always available. While network stability can sometimes be achieved with delayed state information, other performance metrics such as latency may degrade. Thus, instead of simply stabilizing the system, our goal is to design a framework that can mimic arbitrary scheduling algorithms with performance guarantees. A naive approach is to make decisions directly with delayed information, but we show that such methods may lead to poor performance. Instead, we propose the Universal Tracking (UT) algorithm that can mimic the actions of arbitrary scheduling algorithms under observation delay. We rigorously show that the performance gap between UT and the scheduling algorithm being tracked is bounded by constants. Our numerical experiments show that UT significantly outperforms the naive approach in various applications.
\end{abstract}

\section{Introduction} \label{Sec:Intro}

\begin{figure}[htbp]
\centerline{\includegraphics[width=0.98\linewidth]{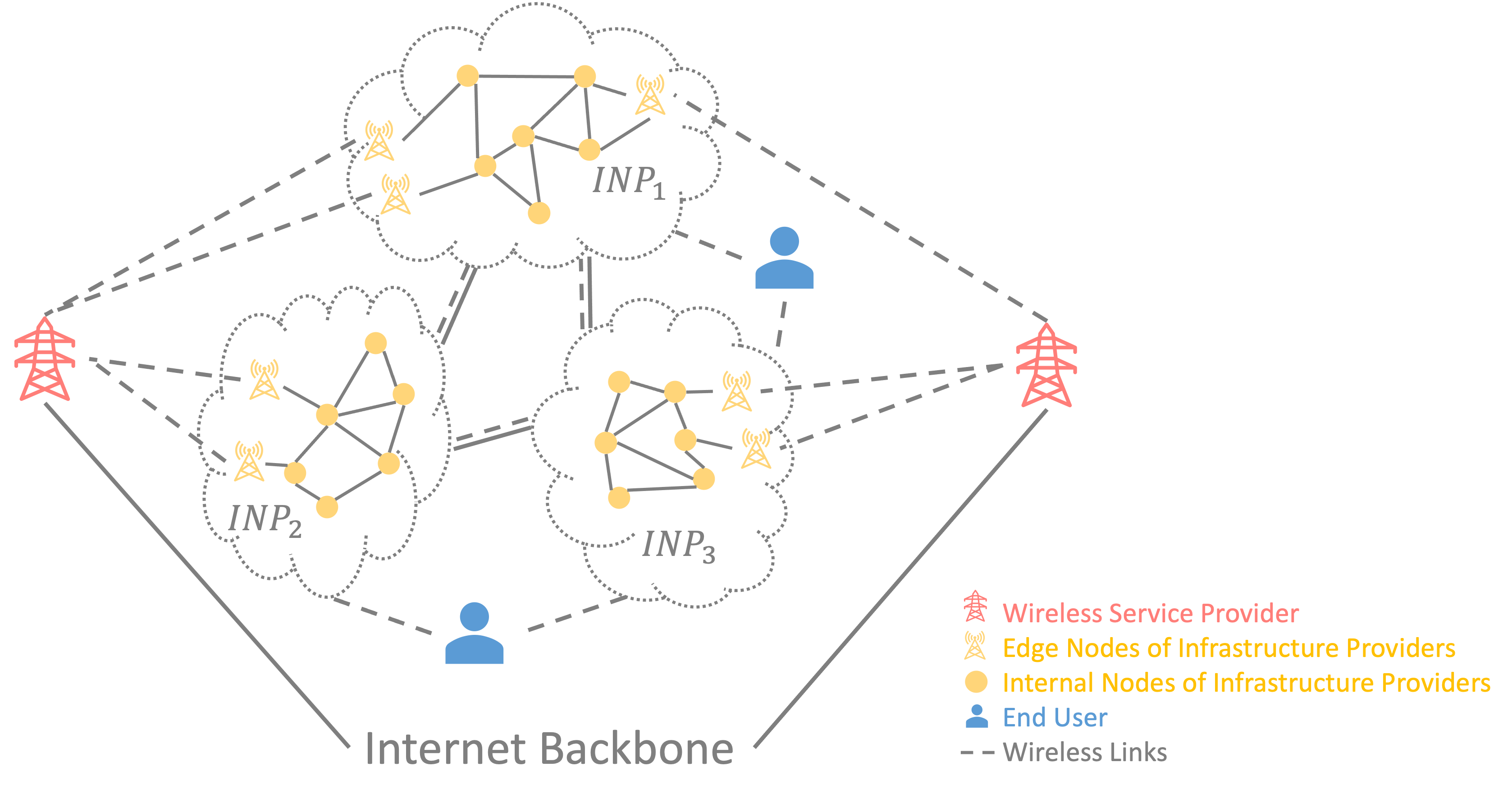}}
\caption{The OTT framework.}
\label{Fig:Intro_Wireless_Model}
\end{figure}

Next generation (NextG) wireless networks have been extensively discussed and studied in recent years. Due to economic concerns, instead of building new infrastructures, an increasing number of NextG wireless service providers (SPs) prefer adapting the over-the-top (OTT) framework \cite{sujata2015impact}. Under the OTT framework, multiple SPs utilize common network infrastructure providers (INPs) to serve end users, as shown in Figure \ref{Fig:Intro_Wireless_Model}. Each end user is connected to one or multiple INPs to send or receive data packets. An INP consists of multiple processing nodes and cannot be directly controlled by the SPs. SPs are connected with the Internet backbone and interact with INPs through edge nodes. The service is bi-directional: in the uplink direction, INPs receive packets from end users and process them, then SPs collect packets (via wireless transmission) from the INPs and send them to the Internet backbone. In the downlink direction, SPs receive data packets from the Internet backbone and dispatch (via wireless transmission) the packets to INPs, the INPs then process the received packets and send them to destined end users. In this paper, we aim to develop a practical scheduling algorithm for general uplink, downlink and bi-directional systems. While there has been enormous amount of work on the problems of wireless scheduling, the issue of delayed state information has received limited attention, and only in the context of specific performance objectives such as stability. In contrast, our approach is to build upon previous work by developing a mechanism to ``track'' the actions of any arbitrary scheduling algorithm.

Most of the existing scheduling algorithms require instantaneous network state information (e.g., backlogs, new external arrivals). However, in practice, queueing, processing and propagation delays may be significant, making it difficult for controllers to obtain fresh network state \cite{goldstein2018synchronization}. It was shown in \cite{liu2021optimal} that throughput optimality can be achieved in general network systems with observation delays. However, many ad-hoc algorithms are designed to optimize other performance metrics such as latency, fairness and power consumption. The performance of such algorithm may suffer if state information is delayed. Therefore, our goal is to design a scheme that can mimic arbitrary scheduling algorithms with desirable performance guarantees.

\begin{figure}
  \centering
  \includegraphics[width=0.75\linewidth]{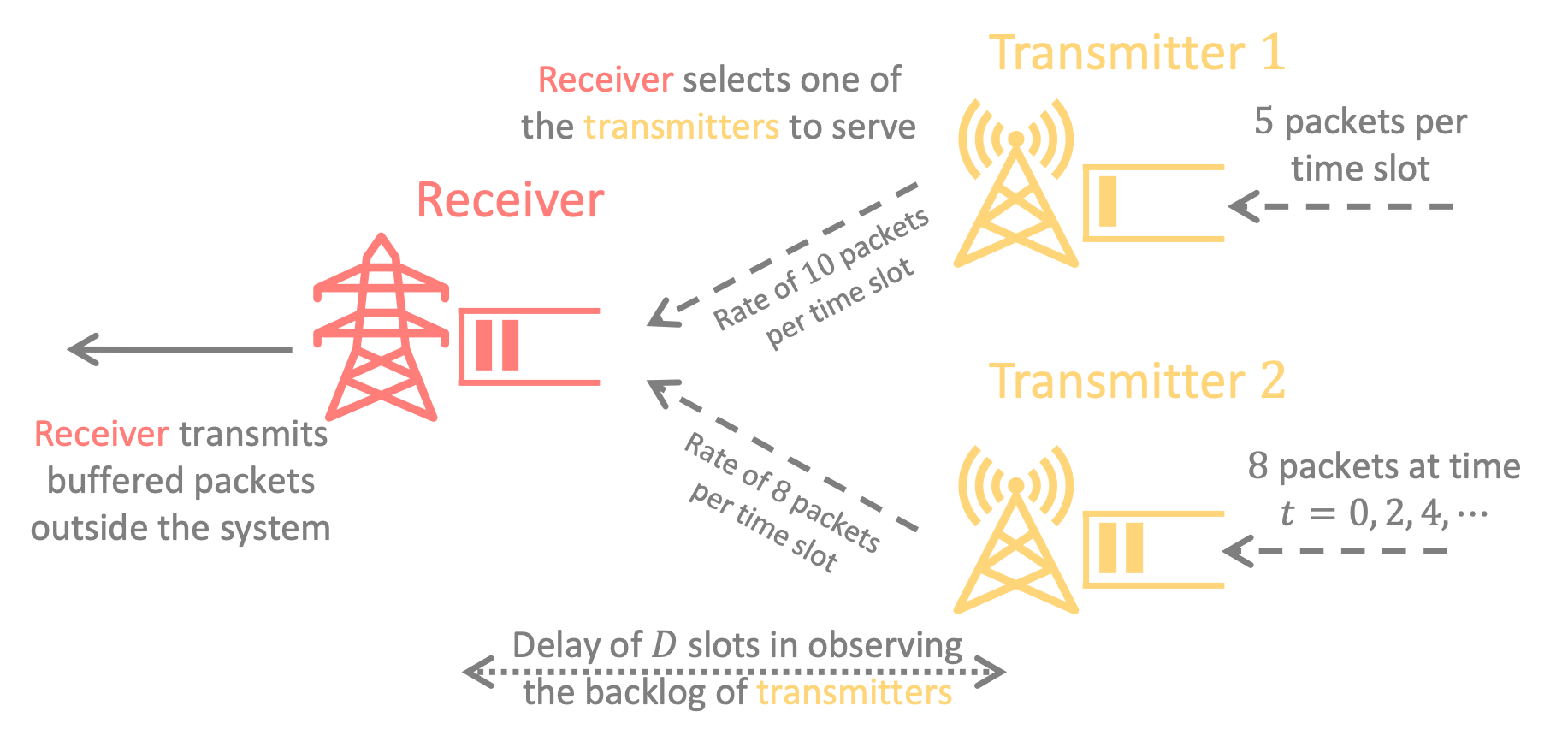}
  \caption{A server allocation system with one receiver and two transmitters.}
  \label{Fig:model_2INPto1SP}
\end{figure}

An intuitive naive approach is to implement the scheduling algorithm based on the delayed network state. However, since the delayed network state may be different from the real time state, the naive approach may have poor performance. Consider a simple server allocation system of one receiver and two transmitters as in Figure \ref{Fig:model_2INPto1SP}. During each time slot, $5$ packets arrive at transmitter $1$, while $8$ packets arrive at transmitter $2$ at time $t = 0, 2, 4, \cdots$. The channel data rates are constantly $10$ and $8$. Due to interference, the receiver can only receive packets from one transmitter in any time slot. If the receiver observes the real time backlogs of the transmitters, we can stabilize the system by always selecting the transmitter with the larger backlog. However, in practice, the receiver only knows the backlog of transmitters $D$ slots ago. Under the naive approach, the receiver always selects the transmitter with larger backlog $D$ slots ago. Since the delayed backlogs may be different from the real time backlogs, the performance suffers. For example, even if $D=1$, it can be shown that the policy with fresh state information has an average backlog of $2.5$ packets, whereas following this naive approach would result in an average backlog of $15.5$ packets. We further propose an example in which the naive approach fails to stabilize the system. The detailed analysis is presented in Section \ref{Sec:compare_INPtoSP}.

The naive approach may fail to follow certain policies and the performance is not guaranteed. In this paper, we propose the Universal Tracking (UT) algorithm that, with delayed observation, can mimic any scheduling policy and obtain provable performance guarantees. We show in Section \ref{Sec:compare_INPtoSP} that UT addresses the stability issues and has significantly better performance than the naive approach.



Numerous scheduling algorithms have been proposed for wireless systems. For instance, for server allocation problems with one server and multiple queues, serving the longest connected queue (LCQ) stabilizes the network \cite{tassiulas1993dynamic} and achieves minimal delay if the system parameters are symmetric \cite{ganti2007optimal}. For load balancing problems with one dispatcher (transmitter) and multiple servers (receivers), join the shortest queue (JSQ) can stabilize the system \cite{foley2001join} and has been widely applied in practice \cite{gupta2007analysis}. For more general server allocation and load balancing problems with multiple transmitters and receivers, the Maximum Weighted Matching (MWM) has been shown to be a stabilizing scheduling scheme \cite{tassiulas1992stability}, yet it is a centralized algorithm and suffers from high computational complexity \cite{papadimitriou1998combinatorial}. An alternative method named Greedy Maximal Matching (GMM) was proposed in \cite{hoepman2004simple}, which can be deployed in a distributed manner and is guaranteed to reach at least $50\%$ of the maximum possible throughput. Moreover, numerous ad-hoc algorithms were designed to optimize other performance metrics including latency \cite{liang2019optimal,liu2019reinforcement}, fairness \cite{mazumdar1991fairness,julian2002qos}, power consumption \cite{baek2004minimizing} and general network utilities \cite{neely2008fairness,xi2007distributed}. 

However, the aforementioned algorithms all require instantaneous network state information and full cooperation among nodes. The limited observability in our model can be captured by an overlay-underlay framework \cite{sitaraman2014overlay}, where some underlay network components are modeled as black boxes and the controllers at overlay can only make decisions with limited underlay information. On the analysis side, numerous works focus on studying the impact imposed by observation delay \cite{yang2006power,johnston2017controller} and operation delay \cite{wang2017heavy}. On the control side, several algorithms that do not require instantaneous underlay state information have been proposed \cite{paschos2014throughput,jones2017overlay,rai2019distributed}. However, the existing overlay-underlay control algorithms only aim at reaching stability. In this paper, our target is to mimic scheduling algorithms with arbitrary objectives.

To design a scheme that mimics general scheduling algorithms, a potential framework is network tracking, which tracks certain variables from the past and make decisions using the tracked information. If the controllers track the actions taken by desired scheduling policies, it is possible to mimic such policies with the tracked information. An application of the tracking framework is to track the actions taken by uncontrollable nodes in overlay-underlay networks with both stochastic dynamics \cite{liang2019optimal} and adversarial dynamics \cite{andrews2004scheduling,andrews2005scheduling,liang2018minimizing,liang2019optimal2}. When the state information is delayed, the work of \cite{liu2021optimal} constructs an emulated system to track the states of the underlay nodes and makes decisions based on the emulated system. However, existing tracking algorithms focus only on stability and cannot mimic general scheduling algorithms.

Therefore, existing scheduling methods either require instantaneous observation of network states, or only guarantee throughput optimality. In this paper, we propose the UT algorithm which, to the best of our knowledge, is the first algorithm that can mimic arbitrary scheduling algorithms with delayed state observations. We rigorously show that the performance gap between UT and the desired policy is upper bounded by the product of delay and total arrival rate. 

We also analyze the naive approach that directly applies the scheduling policy with delayed state information. We propose two examples to show that the naive approach may greatly degrade the performance or even fail to stabilize the system (while UT still achieves stability). Through extensive numerical experiments, we show that UT achieves significant improvements compared with the naive approach under various settings.

The paper is organized as follows. In Section \ref{Sec:INPtoSP}, we study the control for uplink traffic, which can be viewed as a server allocation problem. In Section \ref{Sec:model_INPtoSP} we formulate the problem and introduce notation. Section \ref{Sec:approach_INPtoSP} gives an outline of our approach and presents the details of UT. We rigorously analyze the performance of UT in Section \ref{Sec:analysis_INPtoSP}. In Section \ref{Sec:compare_INPtoSP}, we discuss and analyze the limits of the naive approach with two examples. In Section \ref{Sec:SPtoINP}, we study the control for downlink traffic, which can be viewed as a load balancing problem. Section \ref{Sec:model_SPtoINP}, \ref{Sec:approach_SPtoINP} and \ref{Sec:analysis_SPtoINP} present the model, approach and performance analysis respectively. In Section \ref{Sec:Sim}, we evaluate UT and the naive approach through numerical experiments under dynamic server allocation (Section \ref{Sec:sim_DSA}) and load balancing (Section \ref{Sec:sim_LB}).


\section{Uplink} \label{Sec:INPtoSP}

\begin{figure}
  \centerline{\includegraphics[width=0.75\linewidth]{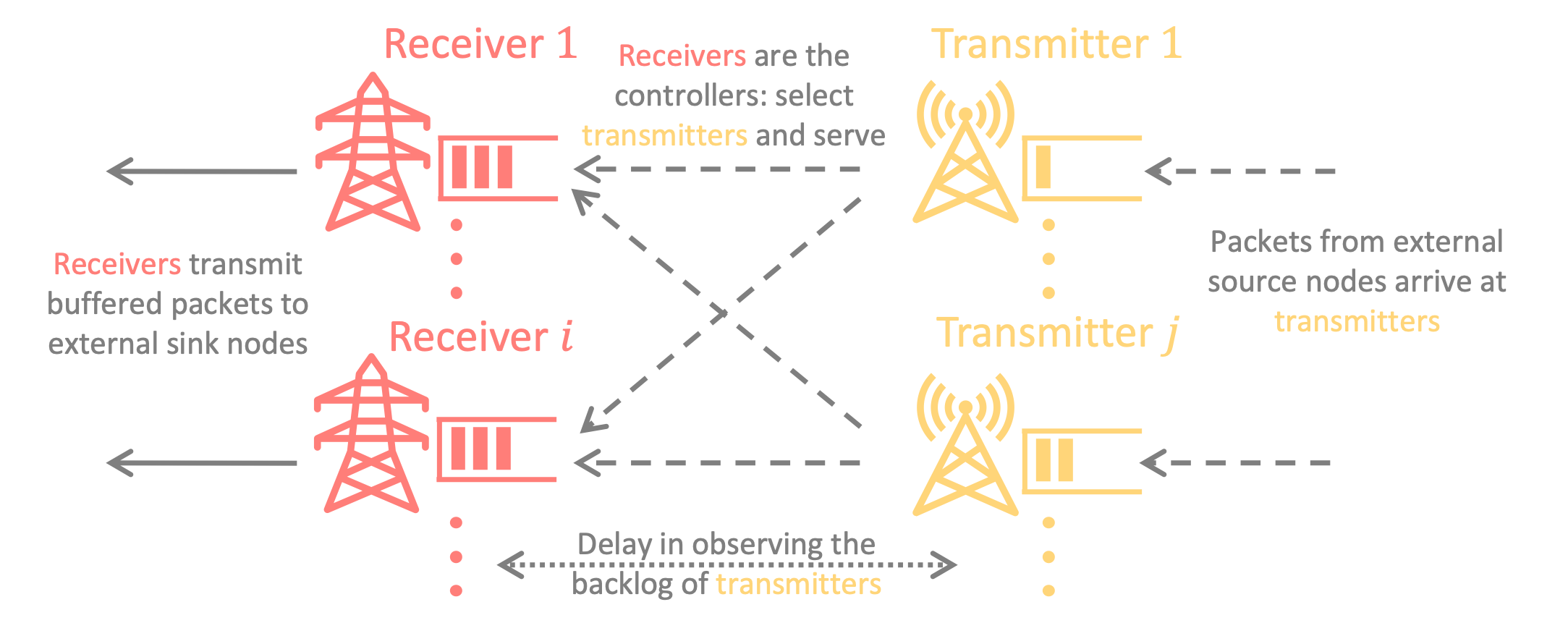}}
  \caption{Uplink system model.}
  \label{Fig:model_INPtoSP}
\end{figure}

We first consider scheduling in the uplink direction. The system is shown in Figure \ref{Fig:model_INPtoSP}. Packets arrive at transmitters from external source nodes. The receivers are controllers and make decisions on selecting transmitters and serving their packets. The receivers then transmit buffered packets to external sink nodes. If we assume that the receivers clear all received packets instantly and have no queue backlogs, the problem can be viewed as a classic server allocation problem \cite{tassiulas1993dynamic}.


\subsection{System Model} \label{Sec:model_INPtoSP}

The sets of receivers and transmitters are denoted by $\mathcal{M}$ and $\mathcal{N}$, respectively. The network has multiple classes of traffic destined to different sink nodes, with the set of classes denoted by $\mathcal{K}$. Each receiver may connect to one or multiple sink nodes, and thus the topology between receivers and sinks nodes can be arbitrary. For simplicity, we use $s$ to denote the aggregation of source nodes and sink nodes outside the system. We assume that the time is slotted and the time horizon is $T$. At the beginning of time slot $t$, transmitter $j$ has $Q_{jk}(t)$ buffered packets of class $k$, and receiver $i$ has $Q_{ik}(t)$ buffered packets of class $k$ that need to be transmitted to a sink node outside the system. Transmitter $j$ receives $A_{jk}(t)$ external packets that are of class $k$. We allow $A_{jk}(t)$ to be non-stochastic and non-stationary over time $t$. The wireless channels between the transmitters and the receivers evolve dynamically under a stationary stochastic process. The current channel data rate from transmitter $j$ to receiver $i$ is $C_{ji}(t)$ and is known to the receivers prior to the decision making process. Receiver $i$ then attempts to serve (receive) $F_{jik}(t)$ packets of class $k$ from transmitter $j$ based on the transmitter state information delayed by $D$ slots. The number of actually served packets $\tilde{F}_{jik}(t)$ may be less than $F_{jik}(t)$ if the buffered packets are less than $F_{jik}(t)$. After receiving packets from the transmitters, receiver $i$ decides to transmit $F_{isk}(t)$ packets to the sink node outside the system. The process is summarized as follows,
\begin{numcases}{}
  Q_{jk}(t+1) = \Big[ Q_{jk}(t) + A_{jk}(t) - \sum_{i \in \mathcal{M}} F_{jik}(t) \Big]^+ \label{Eqn:evolve_real_Qji} \\
  Q_{ik}(t+1) = \Big[ Q_{ik}(t) + \sum_{j \in \mathcal{N}} F_{jik}(t) - F_{isk}(t) \Big]^+ \label{Eqn:evolve_real_Qib}
\end{numcases}
where $[x]^+$ denotes $\max \{0, x \}$.

The problem we aim to solve is, given an uplink traffic scheduling policy $\pi_u$, how to mimic the behavior of applying $\pi_u$ in the ideal system without observation delay. For readers' convenience, we summarize the notations used in this section in Table \ref{Tab:notaion_INPtoSP}.
\begin{table} \caption{Notations of server allocation}
\begin{center}
\begin{tabular}{r p{14cm} }
\toprule
$\mathcal{M}$ & The set of receivers \\
$\mathcal{N}$ & The set of transmitters \\
$\pi_u$ & The uplink scheduling policy to mimic \\
$s$ & Aggregation of the nodes outside the system \\
$T$ & Time horizon \\
$Q_{jk}$ & The number of buffered packets at transmitter $j$ that are of class $k$ \\
$Q_{ik}$ & The number of packets of class $k$ buffered at receiver $i$ that need to be transmitted outside the system \\
$\bm{Q}_u$ & The vector of backlogs of uplink traffic  ($Q_{jk}$'s and $Q_{ik}$'s) \\
$A_{jk}$ & External arrival from end users to transmitter $j$ of class $k$ \\
$\bm{A}_u$ & The vector of arrivals of uplink traffic ($A_{jk}$'s) \\
$C_{ji}$ & Data rate of the wireless link from transmitter $j$ to receiver $i$ \\
$\bm{C}_u$ & The vector of channel data rates of uplink traffic ($C_{ji}$'s) \\
$D$ & Delay in observing transmitters \\
$F_{jik}$ & The number of packets to be transmitted from transmitter $j$ to receiver $i$ \\
$F_{isk}$ & The number of packets to be transmitted to external sinks from receiver $i$ \\
$\bm{F}_u$ & The vector of scheduling actions of uplink traffic ($F_{jik}$'s and $F_{isk}$'s) \\
$\tilde{F}$ & The actual number of transmitted packets, i.e., $\tilde{F} = \min\{F, \text{available packets}\}$ \\
$d_{jk}$ & The time to wait before $Q_{jk}$ becomes empty \\
$d_{ik}$ & The time to wait before $Q_{ik}$ becomes empty \\
$e_{jk}$ & The time elapsed since the last time $Q_{jk}$ was empty \\
\bottomrule
\end{tabular}
\end{center}
\label{Tab:notaion_INPtoSP}
\end{table}


\subsection{Our Approach} \label{Sec:approach_INPtoSP}

Mathematically, the naive approach directly applies $\pi_u$ with the transmitter state information $D$ slots ago, i.e., 
$$
  \bm{F}_u(t) = \pi_u \big( \bm{Q}_u(t-D) + \bm{A}_u(t-D), \bm{C}_u(t) \big) .
$$
We define the actions taken under $\pi_u$ in the ideal system without delay as $\bm{F}^{\pi_u}_u$. It is likely that $\bm{Q}_u(t-D) + \bm{A}_u(t-D)$ significantly differs from $\bm{Q}_u(t) + \bm{A}_u(t)$, which makes the action $\bm{F}_u(t)$ deviate from $\bm{F}^{\pi_u}_u(t)$. Due to the delay, it is impractical to maintain an accurate estimate of $\bm{Q}_u(t) + \bm{A}_u(t)$ and new methods need to be introduced. 

We define the backlog of applying $\pi_u$ to the ideal system with instantaneous network state observation as $\bm{Q}^{\pi_u}_u$. If the receivers can maintain a relatively accurate estimate of $\bm{Q}^{\pi_u}_u(t-D)$ and decide on $\bm{F}_u(t)$ based on the estimate, then $\bm{F}_u(t)$ can mimic the actions in the ideal system $D$ slots ago and approach the ideal performance. However, just having delayed queue information based on $\bm{F}_u$ (as opposed to $\bm{F}^{\pi_u}_u$) is not enough. Thus, we aim to emulate an ideal system based on $\bm{F}^{\pi_u}_u$.

More specifically, our Universal Tracking (UT) algorithm operates in the following manners. During the first $D$ slots, the receivers do not have the state information of the transmitters and just take actions using available information, i.e., 
$$
\bm{F}_u(t) = \pi_u (\text{available information}), \quad 0 \leqslant t \leqslant D-1 .
$$
At time $t = D$, the receivers construct an emulated system with its initial backlogs being the same as the real system, i.e., $\bm{Q}^e_u(0) = \bm{Q}_u(0)$ (we use superscript $e$ to denote variables in the emulated system). 

For time $t \geqslant D$, the receivers compute the action that should be taken under policy $\pi_u$ at time $t-D$, yet with the current channel data rate $\bm{C}_u(t)$, i.e.,
$$
  \bm{F}_u(t) = \pi_u \big( \bm{Q}^e_u(t-D) + \bm{A}_u(t-D), \bm{C}_u(t) \big), \quad t \geqslant D .
$$
The receivers apply $\bm{F}_u(t)$ to the real system, and the backlogs in the real system evolve as \eqref{Eqn:evolve_real_Qji} and \eqref{Eqn:evolve_real_Qib}. We restrict $\bm{F}_u(t)$ not to exceed available packets in the emulated system, which simplifies the evolution in the emulated system as follows.
\begin{numcases}{}
  Q^e_{jk}(t-D+1) = Q^e_{jk}(t-D) + A_{jk}(t-D) - \sum_{i \in \mathcal{M}} F_{jik}(t) \label{Eqn:evolve_emu_Qji} \\
  Q^e_{ik}(t-D+1) = Q^e_{ik}(t-D) + \sum_{j \in \mathcal{N}} F_{jik}(t) - F_{isk}(t) \label{Eqn:evolve_emu_Qib}
\end{numcases}
The receivers then use $\bm{F}_u(t)$ and $\bm{A}_u(t-D)$ to update the emulated system and compute $\bm{Q}_u^e(t-D+1)$ as in \eqref{Eqn:evolve_emu_Qji} and \eqref{Eqn:evolve_emu_Qib}.

The details are presented in Algorithm \ref{Alg:UT_uplink}.

\begin{algorithm}
\caption{The UT algorithm for scheduling uplink traffic} \label{Alg:UT_uplink}
\begin{algorithmic}[1]
  \STATE \textbf{Input: } $\pi_u$, $\bm{Q}_u(0)$
  \STATE Set $\bm{Q}_u^e(0) \leftarrow \bm{Q}_u(0)$
  \FOR{time $t \leftarrow 0, 1, \cdots, D-1$}
    \STATE Observe $\bm{C}_u(t)$
    \STATE Set 
      $$
        \bm{F}_u(t) \leftarrow \pi_u(\text{available information}) 
      $$
    \STATE Apply $\bm{F}_u(t)$ to the real system
  \ENDFOR
  \FOR{time $t \leftarrow D, D+1, \cdots, T-1$}
    \STATE Observe $\bm{C}_u(t)$
    \STATE Observe $\bm{A}_u(t-D)$
    \STATE Set 
      $$
      \bm{F}_u(t) \leftarrow \pi_u \big( \bm{Q}^e_u(t-D) + \bm{A}_u(t-D), \bm{C}_u(t) \big)
      $$
    \STATE Update the emulated system using \eqref{Eqn:evolve_emu_Qji} and \eqref{Eqn:evolve_emu_Qib}
    \STATE Apply $\bm{F}_u(t)$ to the real system
  \ENDFOR
  \STATE \textbf{Output: } a sequence of actions $\{ \bm{F}_u(t) \}_{t=0, 1, \cdots, T-1}$
\end{algorithmic}
\end{algorithm}

We use the example in Figure \ref{Fig:model_2INPto1SP} to illustrate the UT algorithm. We assume the initial backlogs at the transmitters to be $\bm{Q}_u(0) = (0, 0)$. The scheduling policy $\pi_u$ we aim to mimic is to serve the transmitter with larger backlog. During time slot $t = 0$, since the receiver has no state information of the transmitters, we assume it does not serve either transmitter. Therefore, At the beginning of time slot $t = 1$, we have
$$
  \bm{Q}_u(1) = \bm{Q}_u(0) + \bm{A}_u(0) = (5, 8) .
$$

During time slot $t = 1$, the receiver constructs an emulated system with $\bm{Q}^e_u(0) = \bm{Q}_u(0) = (0, 0)$ and observes that $\bm{A}_u(0) = (5, 8)$, which leads to $\bm{Q}^e_u(0) + \bm{A}_u(0) = (5, 8)$. Since transmitter $2$ has larger backlog in the emulated system, the receiver chooses to serve transmitter $2$, i.e., $\bm{F}_u(1) = (0, 8)$. The receiver then applies $\bm{F}_u(1)$ to the real system and uses it to update the emulated system, as follows.
$$
\begin{cases}
  \bm{Q}_u(2) = \bm{Q}_u(1) + \bm{A}_u(1) - \bm{F}_u(1) = (10, 0) \\
  \bm{Q}^e_u(1) = \bm{Q}^e_u(0) + \bm{A}_u(0) - \bm{F}_u(1) = (5, 0)
\end{cases}
.
$$

During time slot $t = 2$, the receiver obtains that $\bm{A}_u(1) = (5, 0)$, and thus $\bm{Q}^e_u(1) + \bm{A}_u(1) = (10, 0)$. Therefore, the receiver chooses to serve transmitter $1$, i.e., $\bm{F}_u(2) = (10, 0)$. The receiver then applies $\bm{F}_u(2)$ to the real system and uses it to update the emulated system, as follows.
$$
\begin{cases}
  \bm{Q}_u(3) = \bm{Q}_u(2) + \bm{A}_u(2) - \bm{F}_u(2) = (5, 8) \\
  \bm{Q}^e_u(2) = \bm{Q}^e_u(1) + \bm{A}_u(1) - \bm{F}_u(2) = (0, 0)
\end{cases}
.
$$

The process repeats afterwards. In Table \ref{Tab:UT_1SP_2INP}, we summarize the process in both the real system and the ideal system without observation delay. As can be seen from the table, the actions taken under UT exactly mimic the ideal actions (delayed by one slot), and the backlogs in the emulated system is exactly the same as the ideal system. Therefore, the emulated system mimics the ideal system. Since the receiver takes actions based on the emulated system, the actions applied to the real system are close to the ideal actions (in a delayed manner). Note that the average backlog using UT is $(5+8+10+0)/2 = 11.5$ packets.
\begin{table}[H]
\setlength{\abovecaptionskip}{0mm}
\setlength{\belowcaptionskip}{3mm}
\centering
\caption{UT Operation for the Uplink System with One receiver and Two transmitters}
\label{Tab:UT_1SP_2INP}
 \begin{tabular}{ccccccc}
  \toprule
  $t$ & $\bm{A}_u(t)$ & Real $\bm{Q}_u(t)$ & Emulated $\bm{Q}_u^e(t)$ & $\bm{F}_u(t)$ & Ideal $\bm{Q}^{\pi_u}_u(t)$ & Ideal $\bm{F}^{\pi_u}_u(t)$ \\
  \midrule
  $0$ & $(5, 8)$ & $(0, 0)$ & $(0, 0)$ & $(0, 0)$ & $(0, 0)$ & $(0, 8)$ \\
  $1$ & $(5, 0)$ & $(5, 8)$ & $(5, 0)$ & $(0, 8)$ & $(5, 0)$ & $(10, 0)$ \\
  $2$ & $(5, 8)$ & $(10, 0)$ & $(0, 0)$ & $(10, 0)$ & $(0, 0)$ & $(0, 8)$ \\
  $3$ & $(5, 0)$ & $(5, 8)$ & $(5, 0)$ & $(0, 8)$ & $(5, 0)$ & $(10, 0)$ \\
  $4$ & $(5, 8)$ & $(10, 0)$ & $(0, 0)$ & $(10, 0)$ & $(0, 0)$ & $(0, 8)$ \\
  $\vdots$ & $\vdots$ & $\vdots$ & $\vdots$ & $\vdots$ & $\vdots$ & $\vdots$ \\
  \bottomrule
\end{tabular}
\end{table}


\subsection{Performance Analysis} \label{Sec:analysis_INPtoSP}

\begin{figure*}[t]
  \centering
  \includegraphics[width=0.98\linewidth]{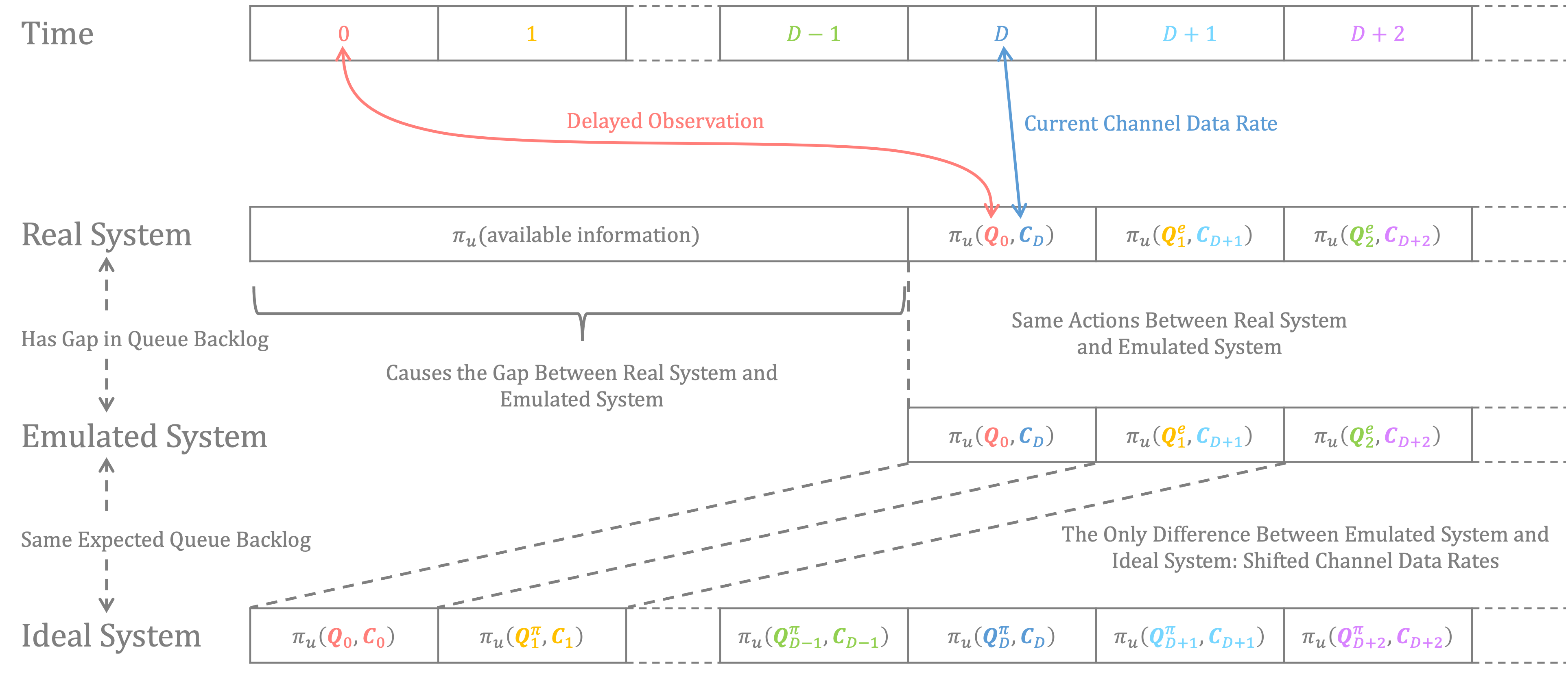}
  \caption{Schemetic illustration of UT.}
  \label{Fig:Approach}
\end{figure*}

As illustrated in Figure \ref{Fig:Approach}, to analyze the performance, we compare the actions taken in three systems: the real system, the emulated system constructed by the receiver and the ideal system which applies $\pi_u$ without observation delay. We first compare the real system and the emulated system. Starting from time $t = D$, both systems take exactly the same actions. However, for the first $D$ time slots, the actions taken by the real system may be arbitrary and thus causing a performance gap. We then compare the emulated system and the ideal system. The emulated system and the ideal system both have instantaneous observation of the transmitters, with the only difference being that the channel data rates are shifted by $D$ slots. We show via stochastic coupling that both systems have the same expected backlogs. Combining the above analysis, we are able to bound the gap between the real system and the ideal system, with the emulated system as a bridge.

We start from the backlogs of packets buffered at transmitters that need to be served by the receivers, i.e., $Q_{jk}$'s. We first analyze a simpler case, where the receivers choose to take no action during the first $D$ slots. We compare the expected average backlog between the real system and the ideal system as in Theorem \ref{Thm:basic_Qji}.
\begin{theorem} \label{Thm:basic_Qji}
    For any arrival sequence $\{ A_{jk}(t) \}_{t = 0, 1, \cdots, T-1}$, any uplink scheduling policy $\pi_u$, each class $k \in \mathcal{K}$, each $i \in \mathcal{M}$ and $j \in \mathcal{N}$, if the receivers do not take actions during the first $D$ slots, we have
    $$
    \mathbb{E} \big[ \bar{Q}_{jk} \big] \leqslant \mathbb{E} \big[ \bar{Q}^{\pi_u}_{jk} \big] + D \cdot \lambda_{jk} ,
    $$
    where $\bar{Q}$ is the average queue backlog over time and $\lambda_{jk}$ is an upper bound for the average $A_{jk}(t)$.
\end{theorem}

\emph{Proof outline: }
We first compare $Q_{jk}$ between the real system and the emulated system. We show that there is always a gap of $\sum_{\tau=t-D}^{t-1} A_{jk}(\tau)$ by induction. We then compare $Q_{jk}$ between the emulated system and the ideal system. We show that if $\bm{C}_u(t)$ in the emulated system are shifted for $D$ slots ahead of the ideal system, both systems share the same backlogs. Since $\bm{C}_u(t)$ have stationary distributions over time $t$, by taking expectations over $\bm{C}(t)$, we show that $Q_{jk}$ has the same expectation between the emulated system and the ideal system. By combining the above analysis, we show that there is always a gap of $\sum_{\tau=t-D}^{t-1} A_{jk}(\tau)$ in the expectation of $Q_{jk}(t)$ between the real system and the ideal system. The detailed proof is given in Appendix \ref{App:basic_Qji}.

We next consider the case where the receivers in the real system take actions based on available information during the first $D$ slots. Intuitively, compared with taking no actions during the first $D$ slots, serving packets during the first $D$ slot should result in reduced gap. However, the analysis is not straightforward since it is possible that starting from $t = D$, the real system has fewer packets and thus the actually served packets in the real system might be less than $F(t)$. Therefore, the analysis in Theorem \ref{Thm:basic_Qji} may not hold and new analysis is required. We derive an upper bound for $Q_{jk}$'s as in Theorem \ref{Thm:UT_Qji}.
\begin{theorem} \label{Thm:UT_Qji}
  For any arrival sequence $\{ A_{jk}(t) \}_{t = 0, 1, \cdots, T-1}$, any uplink scheduling policy $\pi_u$, each class $k \in \mathcal{K}$, each $i \in \mathcal{M}$ and $j \in \mathcal{N}$, we have
  \begin{align*}
      \mathbb{E} \big[ \bar{Q}_{jk} \big] \leqslant & \mathbb{E} \big[ \bar{Q}_{jk}^{\pi_u} \big] + \mathbb{E} \bigg[ \lim_{T \to \infty} \frac{\sum_{t=D}^{T-1} \min\{ D, d_{jk}(t) \} \cdot A_{jk}(t)}{T} \bigg] - \\
      & \mathbb{E} \bigg[ \sum_{\tau = 0}^{D-1} \sum_{i \in \mathcal{M}} \tilde{F}_{jik}(\tau) \cdot \lim_{T \to \infty} \frac{d_{jk}(D) }{T} \bigg] - \\
      & \mathbb{E} \bigg[ \lim_{T \to \infty} \frac{\sum_{t=D}^{T-1} d_{jk}(t) \cdot \mathbbm{1}_{e_{jk}(t) \leqslant D} \cdot A_{ik}(t-D)}{T} \bigg] ,
  \end{align*}
  where $d_{jk}(t)$ denotes the time to wait before $Q_{jk}$ becomes empty, and $e_{jk}(t)$ denotes the time elapsed since the last time $Q_{jk}$ was empty.
\end{theorem}

\emph{Proof outline: }
 We partition the time horizon into intervals during which $Q_{jk} > 0$. Inside each interval, it is guaranteed that $F_{jik}$ does not exceed $Q_{jk} + A_{jk}$ (otherwise $Q_{jk}$ gets emptied). We then show that the accumulated backlog during each interval can be represented as summations of $A_{jk}(t) - A_{jk}(t-D)$, which is upper bounded by the result in Theorem \ref{Thm:UT_Qji} after algebraic operations. The detailed proof is given in Appendix \ref{App:UT_Qji}.

 It is straightforward to see that no matter which actions are taken during the first $D$ slots, the expected average $Q_{jk}$ does not increase: the coefficient of the second term is upper bounded by $D$, and thus the $\mathbb{E} \big[ \bar{Q}_{jk} \big]$ in Theorem \ref{Thm:UT_Qji} is no greater than the result in Theorem \ref{Thm:basic_Qji}.

 We finally compare the backlog of packets buffered at receivers that need to be transmitted outside the system, i.e., $Q_{ik}$'s. We first analyze the case where the receivers choose to take no action during the first $D$ slots, as in Theorem \ref{Thm:basic_Qib}.
 \begin{theorem} \label{Thm:basic_Qib}
  For any arrival sequence $\{ A_{jk}(t) \}_{t = 0, 1, \cdots, T-1}$, any uplink scheduling policy $\pi_u$, each class $k \in \mathcal{K}$ and $i \in \mathcal{M}$, if the receivers do not take actions during the first $D$ slots, we have
  $$
    \mathbb{E} \big[ \bar{Q}_{ik} \big] = \mathbb{E} \big[ \bar{Q}^{\pi}_{ik} \big] .
  $$
\end{theorem}
\emph{Proof outline: }
  We show that $Q_{ik}$ in the real system is always equal to $Q_{ik}$ in the emulated system by induction. We then use the conclusions developed in the proof of Theorem \ref{Thm:basic_Qji} and derive that $Q_{ik}$ has the same expectation between the emulated system and the ideal system. Therefore, the expected $Q_{ik}$ in the real system is equal to the expected $Q_{ik}$ in the ideal system. The detailed proof is given in Appendix \ref{App:basic_Qib}.

We now consider the general case where the receivers take arbitrary actions during the first $D$ slots, and the result is as Theorem \ref{Thm:UT_Qib}.
\begin{theorem} \label{Thm:UT_Qib}
  For any arrival sequence $\{ A_{jk}(t) \}_{t = 0, 1, \cdots, T-1}$, any uplink scheduling policy $\pi_u$, each class $k \in \mathcal{K}$ and $i \in \mathcal{M}$, we have
  $$
    \mathbb{E} \big[ \bar{Q}_{ik} \big] \leqslant \mathbb{E} \big[ \bar{Q}_{ik}^{\pi} \big] + \mathbb{E} \bigg[ \sum_{t = 0}^{D-1} \Big( \sum_{j \in \mathcal{N}} \tilde{F}_{jik}(t) - \tilde{F}_{isk}(t) \Big) \cdot \lim_{T \to \infty} \frac{d_{ik}(D)}{T} \bigg] .
  $$
\end{theorem} 
\emph{Proof outline: }
  Similar to the proof of Theorem \ref{Thm:UT_Qji}, the time horizon is partitioned into intervals during which $Q_{ik} > 0$, and the rest of the proof follows a similar process. The detailed proof is given in Appendix \ref{App:UT_Qib}.

With both Theorem \ref{Thm:UT_Qji} and Theorem \ref{Thm:UT_Qib}, we have the complete performance guarantee for the uplink system as in Theorem \ref{Thm:UT_QjiQib}.
\begin{theorem} \label{Thm:UT_QjiQib}
  For any arrival sequence $\{ \bm{A}_u(t) \}_{t = 0, 1, \cdots, T-1}$ and any uplink scheduling policy $\pi_u$, we have
  $$
  \mathbb{E} \big[ \bar{Q}_u \big] \leqslant \mathbb{E} \big[ \bar{Q}^{\pi_u}_u \big] + D \cdot \sum_{j \in \mathcal{N}, k \in \mathcal{K}} \lambda_{jk} .
  $$
\end{theorem}

The above results show that the gap between the performance of the real system under UT and the ideal system under $\pi_u$ is always upper bounded by the expected number of external arrival during the interval $D$. We emphasize that the uplink scheduling policy $\pi_u$ can be arbitrary, and thus our tracking algorithm has universal applicability.


\subsection{Comparison with Naive Approach} \label{Sec:compare_INPtoSP}

As analyzed in Section \ref{Sec:approach_INPtoSP}, under the UT algorithm, the average backlog of the example in Figure \ref{Fig:model_2INPto1SP} is $11.5$. We use this example to illustrate why the naive approach may degrade the performance compared with UT. Recall that the naive approach directly decides $\bm{F}_u$ with the transmitter state information $D$ slots ago, i.e., 
$$
  \bm{F}_u(t) = \pi_u \big( \bm{Q}_u(t-D) + \bm{A}_u(t-D), \bm{C}_u(t) \big) ,
$$
the process under the naive approach is illustrated in Table \ref{Tab:naive_1SP_2INP}.

\begin{table}[H]
\setlength{\abovecaptionskip}{0mm}
\setlength{\belowcaptionskip}{3mm}
\centering
\caption{Naive Approach Operation for the Uplink System with One receiver and Two transmitters}
\label{Tab:naive_1SP_2INP}
  \begin{tabular}{ccccc}
  \toprule
  $t$ & $\bm{A}_u(t)$ & Real $\bm{Q}_u(t)$ & $\bm{Q}_u(t-1)+\bm{A}_u(t-1)$ & $\bm{F}_u(t)$ \\
  \midrule
  $0$ & $(5, 8)$ & $(0, 0)$ & Undefined & $(0, 0)$ \\
  $1$ & $(5, 0)$ & $(5, 8)$ & $(5, 8)$ & $(0, 8)$ \\
  $2$ & $(5, 8)$ & $(10, 0)$ & $(10, 8)$ & $(10, 0)$ \\
  $3$ & $(5, 0)$ & $(5, 8)$ & $(15, 8)$ & $(10, 0)$ \\
  $4$ & $(5, 8)$ & $(0, 8)$ & $(10, 8)$ & $(10, 0)$ \\
  $5$ & $(5, 0)$ & $(0, 16)$ & $(5, 16)$ & $(0, 8)$ \\
  $6$ & $(5, 8)$ & $(5, 8)$ & $(5, 16)$ & $(0, 8)$ \\
  $7$ & $(5, 0)$ & $(10, 8)$ & $(10, 16)$ & $(0, 8)$ \\
  $8$ & $(5, 8)$ & $(15, 0)$ & $(15, 8)$ & $(10, 0)$ \\
  $9$ & $(5, 0)$ & $(10, 8)$ & $(20, 8)$ & $(10, 0)$ \\
  $10$ & $(5, 8)$ & $(5, 8)$ & $(15, 8)$ & $(10, 0)$ \\
  $11$ & $(5, 0)$ & $(0, 16)$ & $(10, 16)$ & $(0, 8)$ \\
  $12$ & $(5, 8)$ & $(5, 8)$ & $(5, 16)$ & $(0, 8)$ \\
  $\vdots$ & $\vdots$ & $\vdots$ & $\vdots$ & $\vdots$ \\
  \bottomrule
\end{tabular}
\end{table}

From Table \ref{Tab:naive_1SP_2INP}, we observe that at time $t = 12$, the system has the same state as time $t = 6$, which indicates that the system circulates the states between time $t = 6$ and time $t = 11$. Therefore, the average backlog can be calculated by taking the average backlog between time $t = 6$ to time $t = 11$, which gives us $\mathbb{E} \big[ \bar{Q}^{Naive} \big] = 15.5$. Comparing the average backlogs under the two algorithms, we see that even for a simple system and a good uplink scheduling policy $\pi_u$, UT has significant performance improvement compared to the naive approach.

Even worse, the naive approach may fail to stabilize the system. For instance, some flow control algorithms may suspend the service on heavy-loaded links to avoid flooding other functioning links \cite{zhang2002router,geng2022dci}. The naive approach may destabilize the system when mimicking such scheduling algorithms, as illustrated in the following example.

\begin{figure}
  \centering
  \includegraphics[width=0.75\linewidth]{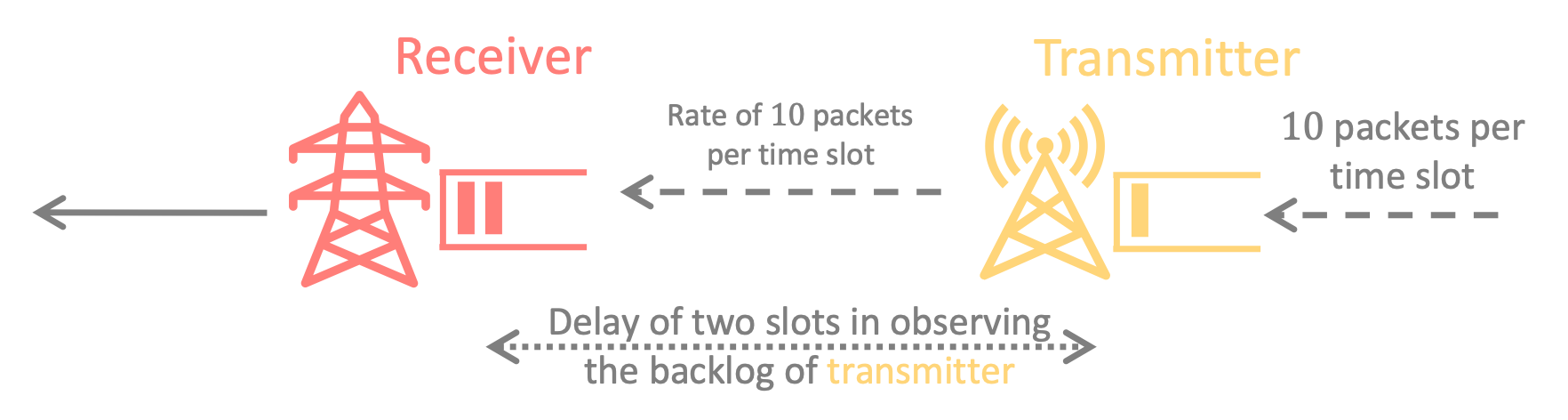}
  \caption{Example of the naive approach failing to stabilize the system.}
  \label{Fig:model_1SP_1INP}
\end{figure}

We consider a toy system with only one receiver and one transmitter as in Figure \ref{Fig:model_1SP_1INP}. The external arrivals to the transmitter are constantly $10$ packets per time slot. The delay for the receiver to observe the transmitter is $D = 2$. The scheduling policy suspends service when the transmitter is congested: the receiver serves $10$ packets when the backlog of the transmitter $Q \leqslant 10$ and does not serve any packet otherwise. All transmitter buffers are initially empty. 

In the ideal system without delay, all external packets get served once they arrive at the transmitter and the queue backlog is always zero. However, in the real system with delay, the naive approach will destabilize the system in the following manner. At time $t = 0$ and $t = 1$, the receiver does not know the state of the transmitter and chooses not to serve, making the transmitter backlog $Q(0) = 10$ and $Q(1) = 20$. At time $t = 2$, the receiver obtains the delayed information that $Q(0) = 10$, and thus serves $10$ packets. However, at time $t = 3$, the receiver observes that $Q(1) = 20$, and thus suspends the service. Similarly, for $t \geqslant 3$, we can show that $Q(t-2) > 10$, which prevents the receiver from serving any packet, and thus destabilizes the system.

On the other hand, it is straightforward to show that by applying the UT algorithm, $Q^e(t) \equiv 0$. Therefore, the receiver always attempts to serve $10$ packets. The average backlog is $20$ and the system is stabilized.

The fundamental reason for the naive approach to have degraded performance is that the observed state information may be distorted by the delay. When the uplink scheduling policy $\pi_u$ is sensitive to backlogs, the performance can degrade significantly. Whereas the UT algorithm, by constructing an emulated system, tracks relatively accurate states under the uplink scheduling policy $\pi_u$ and makes decisions based on them. As we showed in Section \ref{Sec:analysis_INPtoSP}, the UT algorithm can mimic any scheduling policy $\pi$ within a guaranteed gap.


\section{Downlink} \label{Sec:SPtoINP}

\begin{figure}
  \centerline{\includegraphics[width=0.98\linewidth]{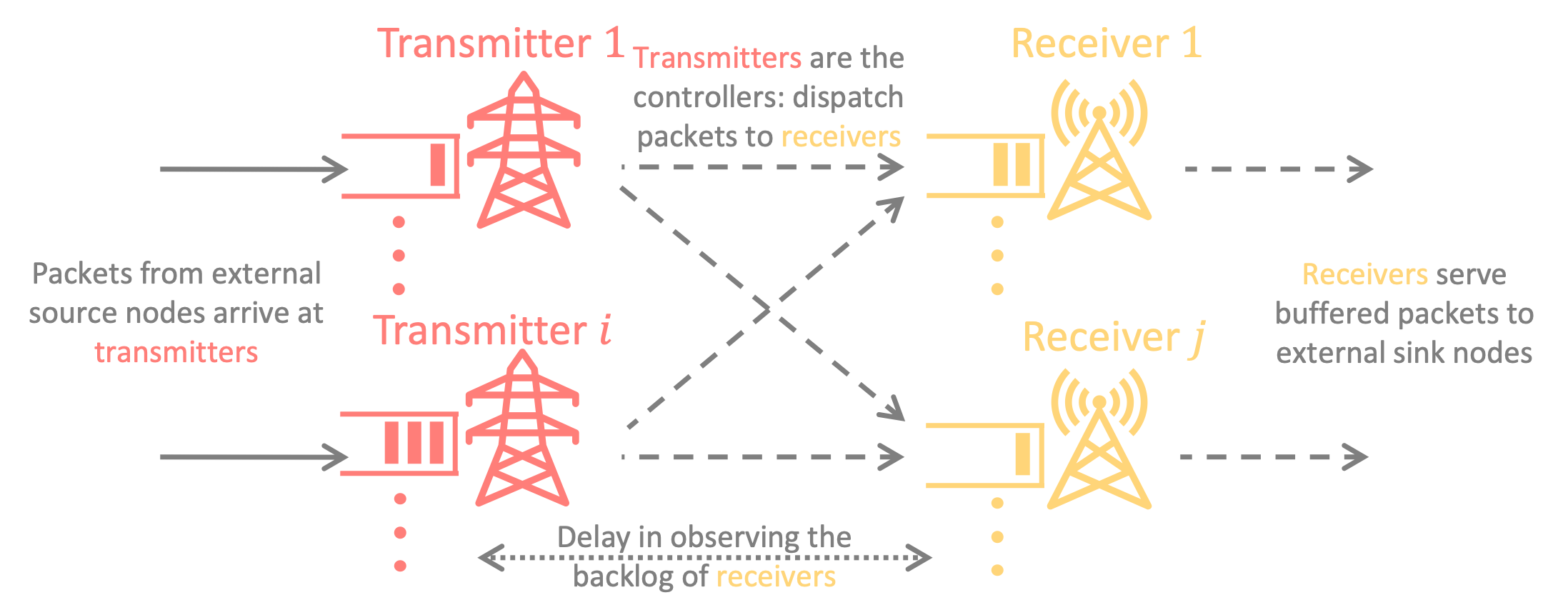}}
  \caption{Downlink system model.}
  \label{Fig:model_SPtoINP}
\end{figure}

We next consider scheduling in the downlink direction. The system is shown in Figure \ref{Fig:model_SPtoINP}. Packets arrive at transmitters from external source nodes. The transmitters are controllers and make decisions to dispatch packets to the receivers. The receivers then transmit buffered packets to external sink nodes according to some unknown policy. Note that the downlink scheduling is significantly different from the uplink scheduling since now the controllers are the transmitters instead of receivers. If we assume that the receivers serve all received packets instantly and have no queue backlogs, the problem can be viewed as a classic load balancing problem \cite{van2018scalable}.


\subsection{System Model} \label{Sec:model_SPtoINP}

The sets of transmitters and receivers are denoted by $\mathcal{M}$ and $\mathcal{N}$, respectively. At the beginning of time slot $t$, transmitter $i$ has $Q_{ik}(t)$ buffered packets destined of class $k$, and receiver $j$ has $Q_{jk}(t)$ buffered packets of class $k$ that need to transmit to external sink nodes. Transmitter $i$ receives $A_{ik}(t)$ external packets of class $k$. We also allow $A_{ik}(t)$ to be non-stochastic and non-stationary over time $t$. The wireless channels from transmitters to receivers evolve dynamically over time following a stationary stochastic process. We assume that transmitter $i$ knows $C_{ij}(t)$, the current channel data rate to receiver $j$, prior to transmission. Transmitter $i$ then decides to transmit $F_{ijk}(t)$ packets to receiver $j$ based on $Q_{jk}(t-D)$, where $D$ is the delay in observing receivers' buffers. After receiving the packets, receiver $j$ attempts to transmit $B_{jsk}(t)$ packets to external sinks. We assume that $B_{jsk}(t)$ follows a stationary stochastic process, but is uncontrollable by the transmitters. The process is summarized as follows.
\begin{numcases}{}
  Q_{ik}(t+1) = \Big[ Q_{ik}(t) + A_{ik}(t) - \sum_{j \in \mathcal{N}} F_{ijk}(t) \Big]^+ \label{Eqn:evolve_real_Qij} \\
  Q_{jk}(t+1) = \Big[ Q_{jk}(t) + \sum_{i \in \mathcal{M}} F_{ijk}(t) - B_{jsk}(t) \Big]^+ \label{Eqn:evolve_real_Qju}
\end{numcases}

Our goal is, given a downlink traffic scheduling policy $\pi_d$, to mimic the behaviors of applying $\pi_d$ in the ideal system without observation delay. For readers' convenience, we summarize the notations used in this section in Table \ref{Tab:notaion_SPtoINP}.
\begin{table}[htbp]\caption{Notations of Downlink Model}
\begin{center}
\begin{tabular}{r p{14cm} }
\toprule
$\pi_d$ & The downlink scheduling policy to mimic \\
$Q_{ik}$ & The number of packets buffered at transmitter $i$ that are destined to receiver $j$ \\
$Q_{jk}$ & The number of packets buffered at receiver $j$ that need to be served \\
$\bm{Q}_d$ & The vector of backlogs of downlink traffic ($Q_{ik}$'s and $Q_{jk}$'s) \\
$A_{ik}$ & External arrival to transmitter $i$ destined to receiver $j$ \\
$\bm{A}_d$ & The vector of downlink traffic arrivals ($A_{ik}$'s) \\
$B_{jsk}$ & Service of receiver $j$ \\
$\bm{B}$ & The vector of $B_{jsk}$'s \\
$\tilde{B}$ & The actual number of transmitted packets, i.e., $\tilde{B} = \min\{B, \text{available packets}\}$ \\
$C_{ji}$ & Data rate of the wireless link from receiver $j$ to transmitter $i$ \\
$\bm{C}_d$ & The vector of channel data rates of downlink traffic ($C_{ij}$'s) \\
$F_{ijk}$ & The number of packets to be transmitted from transmitter $i$ to receiver $j$ \\
$\bm{F}_d$ & The vector of scheduling actions of downlink traffic ($F_{ijk}$'s) \\
$\tilde{F}$ & The actual number of transmitted packets, i.e., $\tilde{F} = \min\{F, \text{available packets}\}$ \\
$d_{ik}$ & The time to wait before $Q_{ik}$ becomes empty \\
$d_{jk}$ & The time to wait before $Q_{jk}$ becomes empty \\
$e_{ik}$ & The time elapsed since the last time $Q_{ik}$ is empty \\
\bottomrule
\end{tabular}
\end{center}
\label{Tab:notaion_SPtoINP}
\end{table}


\subsection{Our Approach} \label{Sec:approach_SPtoINP}

The UT algorithm for downlink traffic is symmetric to the algorithm in Section \ref{Sec:approach_INPtoSP}. We use subscript $d$ to denote vectors for the downlink traffic. The core idea is to let the transmitters maintain an emulated system that estimate the delayed backlogs in the ideal system $\bm{Q}_d^{\pi_d}(t-D)$ and make decisions based on $\bm{Q}_d^{\pi_d}(t-D)$ and $\bm{A}_d(t-D)$. The transmitters update the emulated system with delayed observation of the receiver service $\bm{B}(t-D)$. The evolution in the emulated system is as follows (we also restrict $\bm{F}_d(t)$ not to exceed the available packets in the emulated system).
\begin{numcases}{}
  Q^e_{ik}(t-D+1) = Q^e_{ik}(t-D) + A_{ik}(t-D) - \sum_{j \in \mathcal{N}} F_{ijk}(t) \label{Eqn:evolve_emu_Qij} \\
  Q^e_{jk}(t-D+1) = \Big[ Q^e_{jk}(t-D) + \sum_{i \in \mathcal{M}} F_{ijk}(t) - B_{jsk}(t-D) \Big]^+ \label{Eqn:evolve_emu_Qju}
\end{numcases}

The details are presented in Algorithm \ref{Alg:UT_downlink}.
\begin{algorithm} [H]
\caption{The UT algorithm for scheduling downlink traffic} \label{Alg:UT_downlink}
\begin{algorithmic}[1]
  \STATE \textbf{Input: } $\pi_d$, $\bm{Q}_d(0)$
  \STATE Set $\bm{Q}_d^e(0) \leftarrow \bm{Q}_d(0)$
  \FOR{time $t \leftarrow 0, 1, \cdots, D-1$}
    \STATE Observe $\bm{C}_d(t)$
    \STATE Set 
      $$
        \bm{F}_d(t) \leftarrow \pi_d(\text{available information}) 
      $$
    \STATE Apply $\bm{F}_d(t)$ to the real system
  \ENDFOR
  \FOR{time $t \leftarrow D, D+1, \cdots, T-1$}
    \STATE Observe $\bm{C}_d(t)$
    \STATE Observe $\bm{A}_d(t-D)$ and $\bm{B}(t-D)$
    \STATE Set 
      $$
      \bm{F}_d(t) \leftarrow \pi_d \big( \bm{Q}^e_d(t-D) + \bm{A}_d(t-D), \bm{C}_d(t) \big)
      $$
    \STATE Update the emulated system using \eqref{Eqn:evolve_emu_Qij} and \eqref{Eqn:evolve_emu_Qju}
    \STATE Apply $\bm{F}_d(t)$ to the real system
  \ENDFOR
  \STATE \textbf{Output: } a sequence of actions $\{ \bm{F}_d(t) \}_{t=0, 1, \cdots, T-1}$
\end{algorithmic}
\end{algorithm}


\subsection{Performance Analysis} \label{Sec:analysis_SPtoINP}

The outline of analysis is similar to Section \ref{Sec:analysis_INPtoSP}. By applying similar techniques, we have the following theorems for the backlogs of packets buffered at the transmitters that need to be sent to the receivers, i.e., $Q_{ik}$'s. The proofs are omitted, due to space constraint.
\begin{theorem} \label{Thm:basic_Qij}
  For any arrival sequence $\{ A_{ik}(t) \}_{t = 0, 1, \cdots, T-1}$, any downlink scheduling policy $\pi_d$, each class $k \in \mathcal{K}$, each $i \in \mathcal{M}$ and $j \in \mathcal{N}$, if the transmitters do not take actions during the first $D$ slots, we have
  $$
  \mathbb{E} \big[ \bar{Q}_{ik} \big] \leqslant \mathbb{E} \big[ \bar{Q}^{\pi_d}_{ik} \big] + D \cdot \lambda_{ik} .
  $$
\end{theorem}

\begin{theorem} \label{Thm:UT_Qij}
  For any arrival sequence $\{ A_{ik}(t) \}_{t = 0, 1, \cdots, T-1}$, any downlink scheduling policy $\pi_d$, each class $k \in \mathcal{K}$, each $i \in \mathcal{M}$ and $j \in \mathcal{N}$, we have
  \begin{align*}
      \mathbb{E} \big[ \bar{Q}_{ik} \big] \leqslant & \mathbb{E} \big[ \bar{Q}_{ik}^{\pi_d} \big] + \mathbb{E} \bigg[ \lim_{T \to \infty} \frac{\sum_{t=D}^{T-1} \min\{ D, d_{ik}(t) \} \cdot A_{ik}(t)}{T} \bigg] - \\
      & \mathbb{E} \bigg[ \sum_{\tau = 0}^{D-1} \sum_{j \in \mathcal{N}} \tilde{F}_{ijk}(\tau) \cdot \lim_{T \to \infty} \frac{d_{ik}(D) }{T} \bigg] - \\
      & \mathbb{E} \bigg[ \lim_{T \to \infty} \frac{\sum_{t=D}^{T-1} d_{ik}(t) \cdot \mathbbm{1}_{e_{ik}(t) \leqslant D} \cdot A_{ik}(t-D)}{T} \bigg] .
  \end{align*}
\end{theorem}

We next compare the backlog of packets buffered at receivers that need to be served to end users, i.e., $Q_{jk}$'s, as in Theorem \ref{Thm:basic_Qju}.
\begin{theorem} \label{Thm:basic_Qju}
    For any arrival sequence $\{ A_{ik}(t) \}_{t = 0, 1, \cdots, T-1}$, any downlink scheduling policy $\pi_d$, each class $k \in \mathcal{K}$ and each $j \in \mathcal{N}$, if the transmitters do not take actions during the first $D$ slots, we have
    $$
    \mathbb{E} \big[ \bar{Q}_{jk} \big] \leqslant \mathbb{E} \big[ \bar{Q}^{\pi_d}_{jk} \big] .
    $$
\end{theorem}

\emph{Proof outline: }
  We first compare $Q_{jk}$ between the real system and the emulated system. We show that $Q_{jk}$ in the real system is always no greater than $Q_{jk}$ in the emulated system if the service $\bm{B}(t)$ in the real system are shifted for $D$ slots behind the emulated system. We then compare $Q_{jk}$ between the emulated system and the ideal system. Similar to the proof of Theorem \ref{Thm:basic_Qji}, we show that if $\bm{C}(t)$ in the emulated system are shifted for $D$ slots ahead of the ideal system, both systems share the same backlogs. Since both $\bm{B}(t)$ and $\bm{C}(t)$ have stationary distributions over time $t$, by taking expectations over $\bm{B}(t)$ and $\bm{C}(t)$, we show that the expected $Q_{jk}$ in the real system is always no greater than the expected $Q_{jk}$ in the ideal system. The detailed proof is given in Appendix \ref{App:basic_Qju}.

We now analyze $Q_{jk}$'s when the transmitters take arbitrary actions during the first $D$ slots, as in Theorem \ref{Thm:UT_Qju}.
\begin{theorem} \label{Thm:UT_Qju}
  For any arrival sequence $\{ A_{ik}(t) \}_{t = 0, 1, \cdots, T-1}$, any downlink scheduling policy $\pi_d$, each class $k \in \mathcal{K}$ and each $j \in \mathcal{N}$, we have
  $$
    \mathbb{E} \big[ \bar{Q}_{jk} \big] \leqslant \mathbb{E} \big[ \bar{Q}_{jk}^{\pi_d} \big] + \mathbb{E} \bigg[ \sum_{t = 0}^{D-1} \Big( \sum_{i \in \mathcal{M}} \tilde{F}_{ijk}(t) - \tilde{B}_{jsk}(t) \Big) \cdot \lim_{T \to \infty} \frac{d_{jk}(D)}{T} \bigg] .
    $$
\end{theorem}

\emph{Proof outline: }
  We also let $\bm{B}(t)$ in the real system shifted for $D$ slots behind the emulated system. The time horizon is partitioned into intervals during which $Q_{jk} > 0$, and the rest of the proof follows a similar process to Theorem \ref{Thm:UT_Qib}. The detailed proof is given in Appendix \ref{App:UT_Qju}.

Similarly, for the whole downlink system, we have
\begin{theorem} \label{Thm:UT_QijQju}
  For any arrival sequence $\{ \bm{A}_d(t) \}_{t = 0, 1, \cdots, T-1}$ and any downlink scheduling policy $\pi_d$, we have
  $$
  \mathbb{E} \big[ \bar{Q}_d \big] \leqslant \mathbb{E} \big[ \bar{Q}^{\pi_d}_d \big] + D \cdot \sum_{i \in \mathcal{M}, k \in \mathcal{K}} \lambda_{ik} .
  $$
\end{theorem}

Similar to Theorem \ref{Thm:UT_QjiQib}, Theorem \ref{Thm:UT_QijQju} shows that the gap between the performance of the real system under UT and the ideal system under $\pi_d$ is always upper bounded by the external arrival rates during the delayed durations.


\section{Bi-Directional System} \label{Sec:dual}

The uplink and downlink traffic model in Section \ref{Sec:INPtoSP} and Section \ref{Sec:SPtoINP} are highly symmetric to each other. The servers and the queues in Figure \ref{Fig:model_INPtoSP} correspond to the dispatchers and the servers in Figure \ref{Fig:model_SPtoINP}, respectively. Both the servers in Figure \ref{Fig:model_INPtoSP} and the dispatchers in Figure \ref{Fig:model_SPtoINP} are controllers, with the queues in Figure \ref{Fig:model_INPtoSP} and the servers in Figure \ref{Fig:model_SPtoINP} being uncontrollable and limited in observability. The notations of the model is as Table \ref{Tab:notaion_dual}.
\begin{table}[htbp]\caption{Notations of Bi-Directional Model}
\begin{center}
\begin{tabular}{r p{14cm} }
\toprule
$\pi$ & The scheduling policy to mimic \\
$\bm{Q}$ & The vector of all backlogs \\
$\bm{A}$ & The vector of arrivals in both directions ($A_{jk}$'s and $A_{ik}$'s) \\
$\bm{B}$ & The vector of $B_{jsk}$'s \\
$\bm{C}$ & The vector of channel data rates in both directions ($C_{ji}$'s and $C_{ij}$'s) \\
$D$ & Delay in observing INPs \\
$\bm{F}$ & The vector of scheduling actions in both directions ($F_{jik}$'s, $F_{ijk}$'s and $F_{isk}$'s) \\
\bottomrule
\end{tabular}
\end{center}
\label{Tab:notaion_dual}
\end{table}

Our approach for the bi-directional system combines Section \ref{Sec:approach_INPtoSP} and Section \ref{Sec:approach_SPtoINP} by constructing a bi-directional emulated system and use $\bm{A}(t-D)$ and $\bm{B}(t-D)$ to update it. The details are presented in Algorithm \ref{Alg:UT_dual}.
\begin{algorithm} [H]
\caption{The UT Algorithm for Scheduling Bi-Directional Traffic} \label{Alg:UT_dual}
\begin{algorithmic}[1]
  \STATE \textbf{Input: } $\pi$, $\bm{Q}(0)$
  \STATE Set $\bm{Q}^e(0) \leftarrow \bm{Q}(0)$
  \FOR{time $t \leftarrow 0, 1, \cdots, D-1$}
    \STATE Observe $\bm{C}(t)$
    \STATE Set 
      $$
        \bm{F}(t) \leftarrow \pi(\text{available information}) 
      $$
    \STATE Apply $\bm{F}(t)$ to the real system
  \ENDFOR
  \FOR{time $t \leftarrow D, D+1, \cdots, T-1$}
    \STATE Observe $\bm{C}(t)$
    \STATE Observe $\bm{A}(t-D)$ and $\bm{B}(t-D)$
    \STATE Set 
      $$
      \bm{F}(t) \leftarrow \pi \big( \bm{Q}^e(t-D) + \bm{A}(t-D), \bm{C}(t) \big)
      $$
    \STATE Update the emulated system using \eqref{Eqn:evolve_emu_Qji}, \eqref{Eqn:evolve_emu_Qib}, \eqref{Eqn:evolve_emu_Qij} and \eqref{Eqn:evolve_emu_Qju}
    \STATE Apply $\bm{F}(t)$ to the real system
  \ENDFOR
  \STATE \textbf{Output: } a sequence of actions $\{ \bm{F}(t) \}_{t=0, 1, \cdots, T-1}$
\end{algorithmic}
\end{algorithm}

The performance analysis conducted in Section \ref{Sec:analysis_INPtoSP} and Section \ref{Sec:analysis_SPtoINP} still hold and we have the following upper bound for the average total backlog by summing up the results in Theorem \ref{Thm:UT_QjiQib} and \ref{Thm:UT_QijQju}.
\begin{theorem} \label{Thm:UT_dual}
  For any arrival sequence $\{ \bm{A}(t) \}_{t = 0, 1, \cdots, T-1}$ and any bi-directional scheduling policy $\pi$, we have
  $$
  \mathbb{E} \big[ \bar{Q} \big] \leqslant \mathbb{E} \big[ \bar{Q}^{\pi} \big] + D \cdot \sum_{i \in \mathcal{M}, j \in \mathcal{N}, k \in \mathcal{K}} \big( \lambda_{ik} + \lambda_{jk} \big) .
  $$
\end{theorem}


\section{Numerical Experiments} \label{Sec:Sim}

In this section, we evaluate the performance of UT by conducting numerical experiments. In each case, we compare the performance among classic scheduling algorithms (assuming instantaneous observation), UT and the naive approach (also taking actions during the first $D$ slots).


\subsection{Uplink} \label{Sec:sim_DSA}

\begin{figure}
    \centering
    \includegraphics[width=0.6\linewidth]{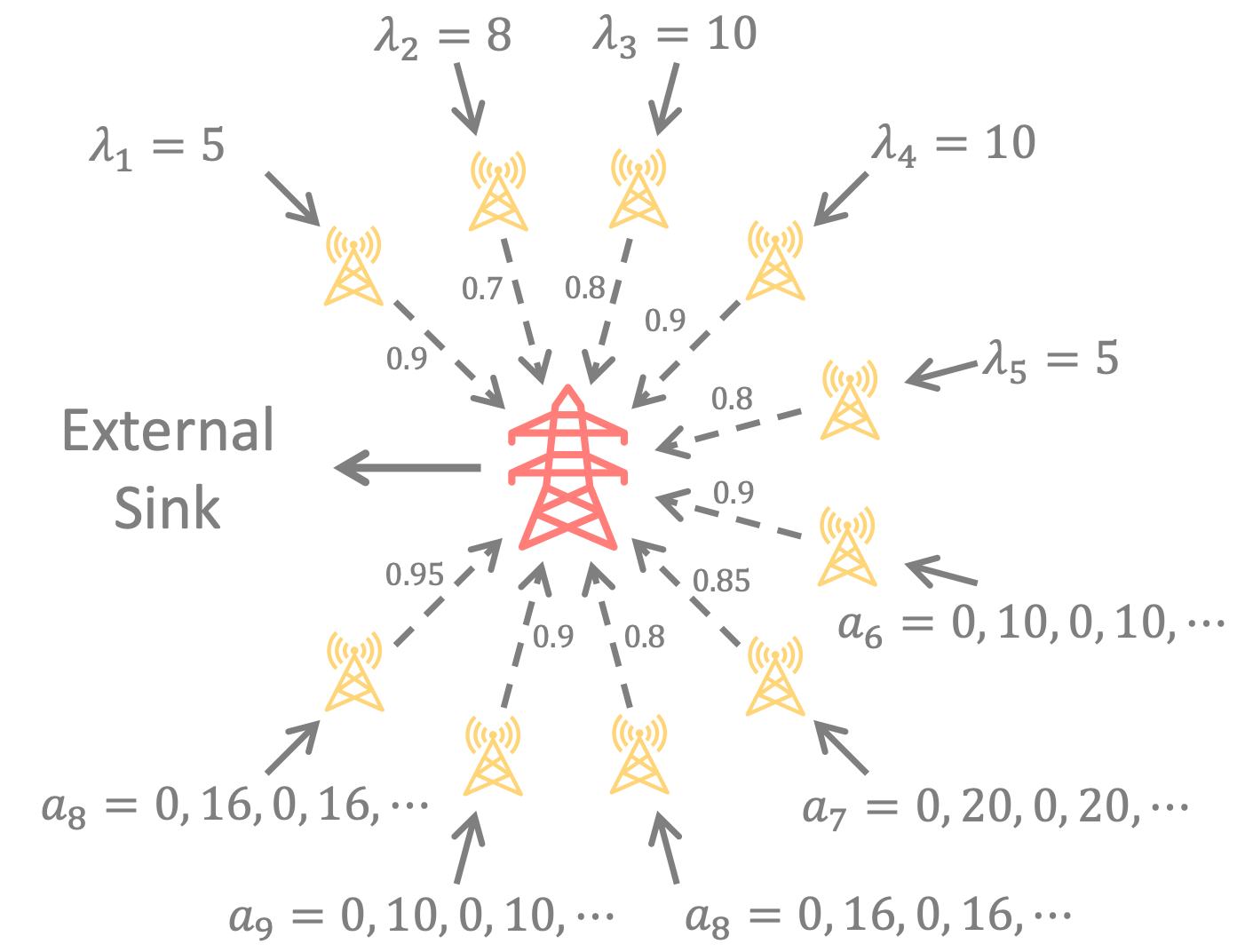}
    \caption{Dynamic server allocation system model.}
    \label{Fig:DSA_model}
\end{figure}

We consider a system of uplink direction with one receiver and ten transmitters. Only one transmit channel can be activated during a time slot. At the beginning of each time slot, external packets arrive at receivers. The receiver then observes the channel data rates and select a transmitter to serve its buffered packets. Such systems are called dynamic server allocation systems. It is shown in \cite{tassiulas1993dynamic} that if the channel data rates are binary, the throughput optimality is obtained by always selecting the connected transmitter with the longest queue (LCQ). From simulation, we find that LCQ performs well for more general settings, and is thus an ideal uplink scheduling policy to mimic. However, LCQ requires instantaneous observation of the transmitter backlogs, which may be unrealistic in practice. The system parameters are shown in Figure \ref{Fig:DSA_model}. 

\begin{figure}
    \centering
    \includegraphics[width=0.7\linewidth]{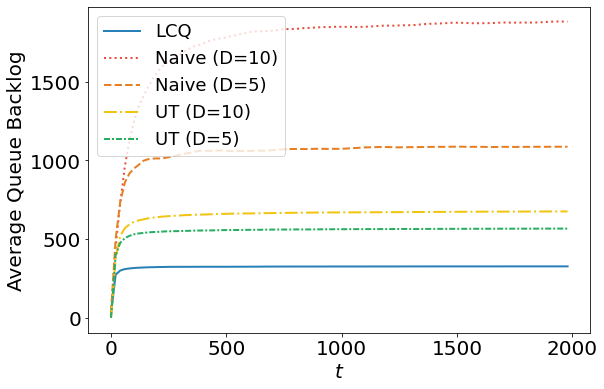}
    \caption{Simulation result for dynamic receiver allocation problem.}
    \label{Fig:DSA_result}
\end{figure}

The receiver obtains the backlog information cyclically from transmitter $1$ to transmitter $10$. The external arrival processes to the transmitters are mixed. The arrival to transmitter $i = 1, 2, 3, 4, 5$ is of Poisson distributions with rate $\lambda_i$'s. While for transmitter $i = 6, 7, 8, 9, 10$, the arrival is not stationary over time, but according to the annotated arrival sequences. The receiver and the transmitters are connected with the annotated probabilities, and the channel data rates are $100$ once connected. We assume that the data rates between the receiver and the sink nodes are large enough so that all packets are immediately cleared once they arrive at the receiver. The scheduling policy $\pi$ we track is the LCQ policy. In both UT and the naive approaches, during the first ten slots, the receiver decides its action based on the available information (i.e., queue backlogs of some transmitters). The results are shown in Figure \ref{Fig:DSA_result}.

Since we assume instantaneous observation when implementing LCQ, its performance serves as a lower bound and may not be achievable in the actual system. From the simulation results, we can see that higher observation delay downgrades the performance for both UT and the naive approach. However, even when $D = 10$, UT significantly outperforms the naive approach.


\subsection{Downlink} \label{Sec:sim_LB}

\begin{figure}
    \centering
    \includegraphics[width=0.6\linewidth]{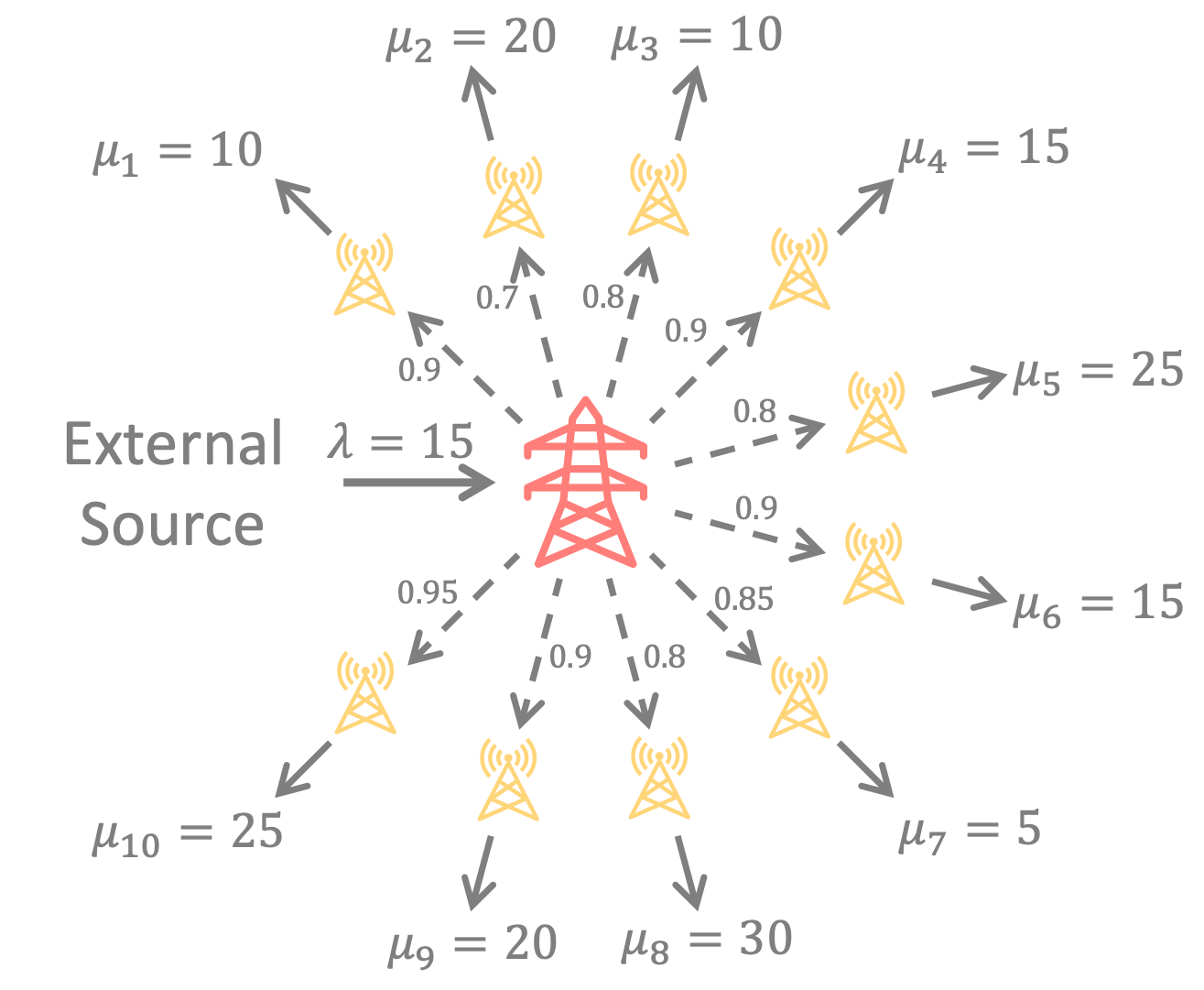}
    \caption{Load balancing system model.}
    \label{Fig:LB_model}
\end{figure}

We consider a system similar to Figure \ref{Fig:DSA_model}, but the data packets now only flow in the downlink direction: external packets arrive to transmitters, the transmitters transmit packets to the receivers, and the receivers serve the packets. During each time slot, the transmitter needs to select one of the receivers to dispatch buffered packets. Such scheduling problems are called load balancing problems, and a throughput optimal policy is known to be joining the shortest queue (JSQ) \cite{van2018scalable}. We consider a load balancing problem as shown in Figure \ref{Fig:LB_model}.

\begin{figure}
    \centering
    \includegraphics[width=0.7\linewidth]{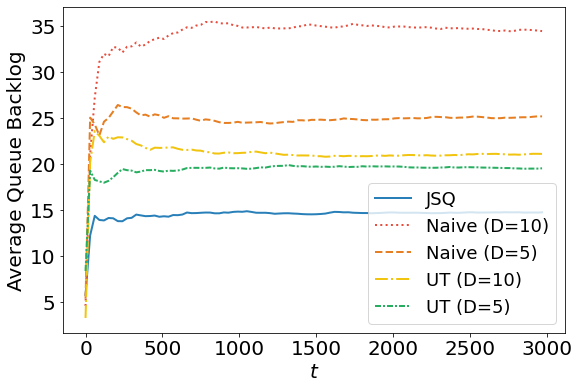}
    \caption{Simulation result for load balancing problem.}
    \label{Fig:LB_result}
\end{figure}

The transmitter obtains the state information of the receivers in a similar manner to Section \ref{Sec:sim_DSA}. The external arrivals to the transmitter are Poisson distributed with rate $\lambda = 15$. The transmitter and the receivers are connected with annotated probabilities, and the channel data rates are $100$ once connected. The service process from receiver $j$ to end users is of uniform distribution with the annotated rate $\mu_j$. The results are shown in Figure \ref{Fig:LB_result}.

Similar to the simulation in Section \ref{Sec:sim_DSA}, JSQ has instantaneous observation of the receiver queues and thus only serves as a lower bound for comparison. The results also show that smaller observation delay help improve the performance, and UT significantly outperforms the naive approach.


\subsection{Bi-Directional System} \label{Sec:sim_dual}

\begin{figure}[H]
    \centering
    \includegraphics[width=0.6\linewidth]{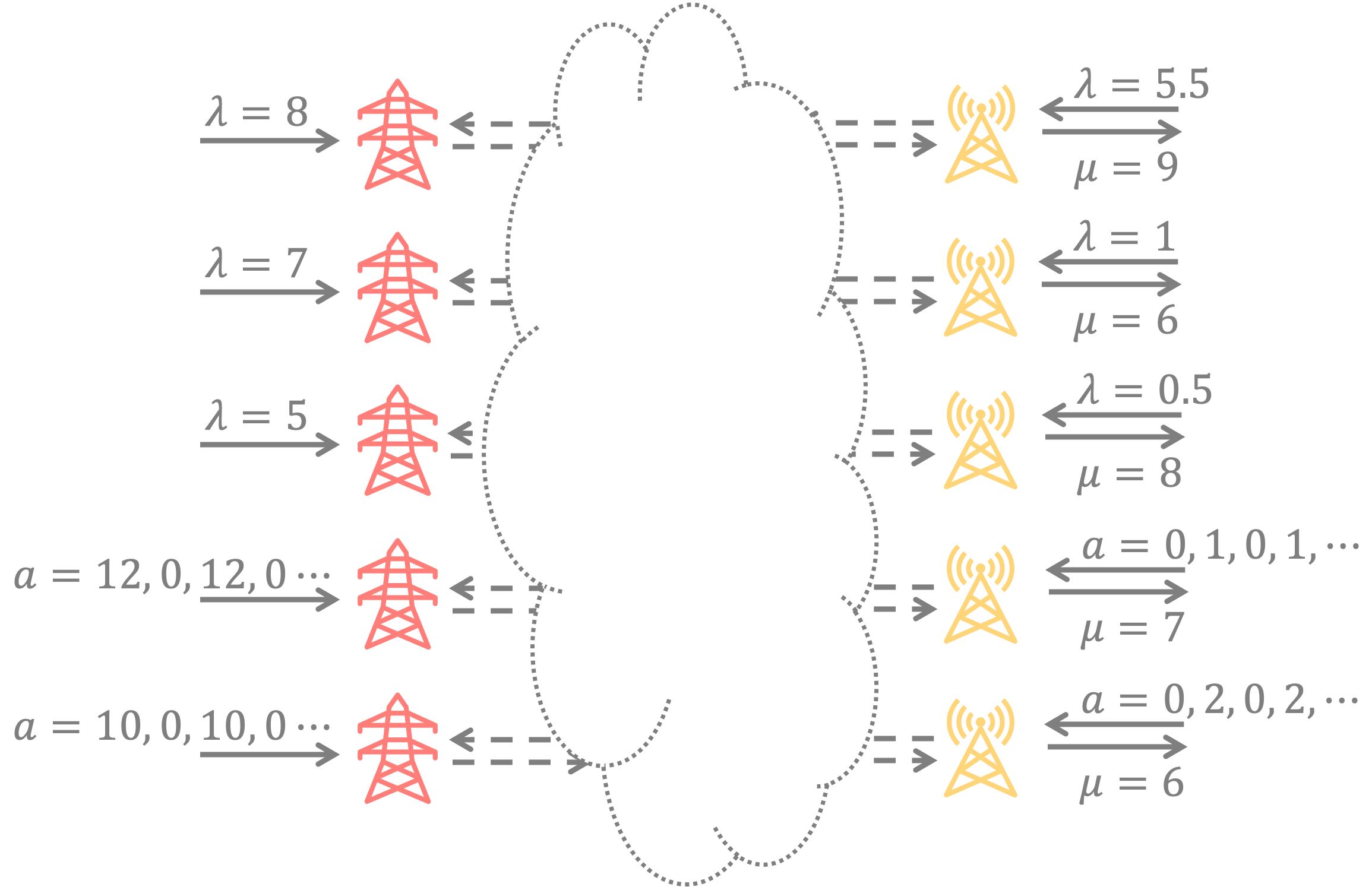}
    \caption{Bi-directional communication system model.}
    \label{Fig:dual_model}
\end{figure}

We turn to a bi-directional system that combines both uplink and downlink traffic. For conciseness, we use the terms introduced in Section \ref{Sec:Intro} to denote nodes in the bi-directional system. We denote the nodes that are receivers in the uplink direction and are transmitters in the downlink direction as SPs, and the nodes that are transmitters in the uplink direction and are receivers in the downlink direction as INPs. Thus, the two directions in the system are: from right source to INPs to SPs and to left sinks, and from left source to SPs to INPs to right sinks. The system consists of five SPs and five INPs, with each SP reachable by each INP, and vice versa. We assume the bi-directional flows are operated independently, i.e., the uplink traffic is scheduled during the first half time slot, and the downlink traffic is scheduled during the second half time slot. To achieve throughput optimality in both directions, the SPs need to pair up with INPs such that the sum of service rate and backlog difference is maximized \cite{tassiulas1992stability}. Related algorithms include MWM and GMM, as discussed in Section \ref{Sec:Intro}. MWM is guaranteed to find the optimal pairing-ups but suffers from high time complexity. GMM has lower time complexity and can be implemented in a distributed manner, but may get stuck at local optimum. The system is as Figure \ref{Fig:dual_model}.

\begin{figure}[H]
    \centering
    \includegraphics[width=0.7\linewidth]{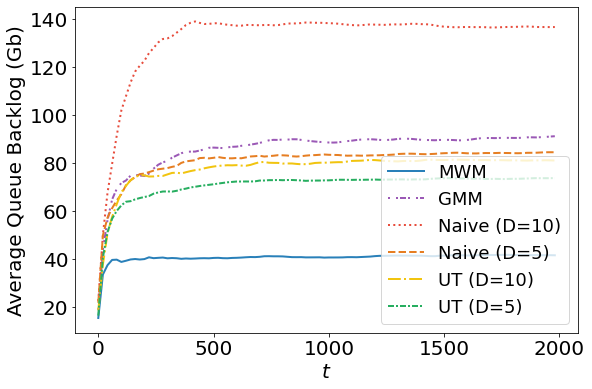}
    \caption{Simulation result for bi-directional communication problem.}
    \label{Fig:dual_result}
\end{figure}

External arrivals to SPs and INPs are either of Poisson distribution with annotated rates or the annotated non-stationary sequences. To simulate the real environment in wireless communication, we adapt the channel rate distribution as in Figure 14 in \cite{yenihayat2016analytical}. The service processes from INPs to end users are of uniform distributions with the annotated rates. All packets arrive from the INPs to the SPs are immediately transmitted to the left sinks. The results are shown in Figure \ref{Fig:dual_result}. 

Both MWM and GMM are conducted with instantaneous observation and only serve for comparisons. The scheduling policy to mimic is MWM. From the simulation results, even if UT is applied with relatively high delay, it outperforms GMM with instantaneous observation. Given the same observation delay, UT still has significant performance improvement compared with the naive approach.


\section{Conclusion} \label{Sec:Conclusion}

In this paper, we focus on mimicking arbitrary scheduling policies in wireless networks with delayed state information. We propose the UT algorithm that can mimic any scheduling policy, and show that the gap between UT and the desired policy (assuming instantaneous observation) is upper bounded by a constant. Bi-directional systems with independent uplink and downlink traffic also fit into our analysis. Numerical experiments validate our conclusion and show that UT has significantly better performance under various settings.

For future works, a potential direction is to analyze the system with adversarial dynamics (e.g., the external arrivals evolve dynamically according to our actions to maximize attack). Another possible direction is to coordinate with the policies inside the INPs and extend the framework into multi-hop settings.


\section*{Acknowledgement}

This work was supported by NSF grants CNS-1907905 and CNS-1735463 and by ONR grant N00014-20-1-2119.

\appendix
\appendixpage


\section{Proof of Theorem \ref{Thm:basic_Qji}} \label{App:basic_Qji}

We first compare $Q_{jk}(t)$ and $Q^e_{jk}(t-D)$. When $t = D$, since the servers take no action during time $0 \leqslant t \leqslant D-1$, we simply have
\begin{equation} \label{Eqn:basic_real_emu_Qji_D}
  Q_{jk}(D) = Q^e_{jk}(0) + \sum_{\tau=0}^{D-1} A_{jk}(\tau) .
\end{equation}

Suppose when $t = w$ where $w \geqslant D$, we have
\begin{equation} \label{Eqn:basic_real_emu_Qji_k}
  Q_{jk}(w) = Q^e_{jk}(w-D) + \sum_{\tau=w-D}^{w-1} A_{jk}(\tau) .
\end{equation}

When $t = k+1$, in the emulated system, by \eqref{Eqn:evolve_emu_Qji}, we have
\begin{equation}  \label{Eqn:basic_real_emu_Qji_kp1}
    Q^e_{jk}(w-D+1) = Q^e_{jk}(w-D) + A_{jk}(w-D) - \sum_{i \in \mathcal{M}} F_{jik}(w) .
\end{equation}

Meanwhile, in the real system, by \eqref{Eqn:evolve_real_Qji}, we have
\begin{align}
    Q_{jk}(w+1) = & \max \big\{ 0, Q_{jk}(w) + A_{jk}(w) - \sum_{i \in \mathcal{M}} F_{jik}(w) \big\} \nonumber \\
    = & \max \big\{ 0, Q^e_{jk}(w-D) + \sum_{\tau=k-D}^{k-1} A_{jk}(\tau) + A_{jk}(w) - \sum_{i \in \mathcal{M}} F_{jik}(w) \big\} \nonumber \\
    = & \max \big\{ 0, Q^e_{jk}(w-D) + A_{jk}(w-D) + \sum_{\tau=k-D+1}^{k} A_{jk}(\tau) - \sum_{i \in \mathcal{M}} F_{jik}(w) \big\} \nonumber \\
    = & \max \big\{ 0, Q^e_{jk}(w-D+1) + \sum_{\tau=k-D+1}^{k} A_{jk}(\tau) \big\} \nonumber \\
    = & Q^e_{jk}(w-D+1) + \sum_{\tau=k-D+1}^{k} A_{jk}(\tau) \label{Eqn:basic_real_emu_Qji_kp1_2} ,
\end{align}
where the second equation holds by inserting \eqref{Eqn:basic_real_emu_Qji_k} and the fourth equation holds by applying \eqref{Eqn:basic_real_emu_Qji_kp1}.

By combining \eqref{Eqn:basic_real_emu_Qji_D}, \eqref{Eqn:basic_real_emu_Qji_k} and \eqref{Eqn:basic_real_emu_Qji_kp1_2}, we have that, by induction, for $t \geqslant D$,
\begin{equation}  \label{Eqn:basic_real_emu_Qji_t}
  Q_{jk}(t) = Q^e_{jk}(t-D) + \sum_{\tau=k-D}^{t-1} A_{jk}(\tau) .
\end{equation}

We then compare $Q^e_{jk}(t)$ and $Q^{\pi_u}_{jk}(t)$. We use $\bm{C}_{t_1:t_2}$ to denote the vector of $\bm{C}_u(t)$ from time $t_1$ to $t_2$. Equations from \eqref{Eqn:basic_emu_ideal_all_0} to \eqref{Eqn:basic_emu_ideal_all_kp1} are all under the condition that $\bm{C}_{D:T-1} = \bm{c}_{0:T-D-1}$ for the emulated system, $\bm{C}_{0:T-D-1} = \bm{c}_{0:T-D-1}$ for the ideal system and $\bm{B}_{0:T-D-1} = \bm{b}_{0:T-D-1}$ for both systems.

Since both systems have the same initial state, we have
\begin{equation}  \label{Eqn:basic_emu_ideal_all_0}
  \bm{Q}_u^e(0) = \bm{Q}_u^{\pi_u}(0) .
\end{equation}

Suppose when $t = k$ where $k \leqslant T-D-2$, we have 
\begin{equation} \label{Eqn:basic_emu_ideal_all_k}
  \bm{Q}_u^e(w) = \bm{Q}_u^{\pi_u}(w) .
\end{equation}

When $t = k+1$, in the emulated system, by \eqref{Eqn:evolve_emu_Qji}, we have
\begin{align}
  Q^e_{jk}(w+1) = & Q^e_{jk}(w) + A_{jk}(w) - \sum_{i \in \mathcal{M}} F_{jik}(w+D) \nonumber \\
  = & Q^e_{jk}(w) + A_{jk}(w) - \pi_u \big( \bm{Q}_u^e(w) + \bm{A}_u(w), \bm{C}_u(w+D) \big) \nonumber \\
  = & Q^e_{jk}(w) + A_{jk}(w) - \pi_u \big( \bm{Q}_u^e(w) + \bm{A}_u(w), \bm{c}_w \big) \label{Eqn:basic_emu_ideal_Qji_kp1} .
\end{align}

Meanwhile, in the ideal system, we have
\begin{align}
  Q^{\pi_u}_{jk}(w+1) = & \max \big\{ 0, Q^{\pi_u}_{jk}(w) + A_{jk}(w) - \sum_{i \in \mathcal{M}} F_{jik}(w+D) \big\} \nonumber \\
  = & \max \big\{ 0, Q^{\pi_u}_{jk}(w) + A_{jk}(w) - \pi_u \big( \bm{Q}_u^{\pi_u}(w) + \bm{A}_u(w), \bm{C}_u(w) \big) \big\} \nonumber \\
  = & \max \big\{ 0, Q^{\pi_u}_{jk}(w) + A_{jk}(w) - \pi_u \big( \bm{Q}_u^{\pi_u}(w) + \bm{A}_u(w), \bm{c}_w \big) \big\} \label{Eqn:basic_emu_ideal_Qji_kp1_2} .
\end{align}

Combining \eqref{Eqn:basic_emu_ideal_all_k}, \eqref{Eqn:basic_emu_ideal_Qji_kp1} and \eqref{Eqn:basic_emu_ideal_Qji_kp1_2}, we show that $Q^e_{jk}(w+1) = Q^{\pi_u}_{jk}(w+1)$. Similar analysis also holds for $Q_{ik}$, which gives us
\begin{equation} \label{Eqn:basic_emu_ideal_all_kp1}
  \bm{Q}_u^e(w+1) = \bm{Q}_u^{\pi_u}(w+1) .
\end{equation}

By combining \eqref{Eqn:basic_emu_ideal_all_0}, \eqref{Eqn:basic_emu_ideal_all_k} and \eqref{Eqn:basic_emu_ideal_all_kp1}, we have that, by induction, for $t \leqslant T-D-1$,
\begin{align}
    & \mathbb{E} \big[ \bm{Q}_u^e(t) \mid \bm{B}_{0:T-D-1} = \bm{b}_{0:T-D-1}, \bm{C}_{D:T-1} = \bm{c}_{0:T-D-1} \big] \nonumber \\
    = & \mathbb{E} \big[ \bm{Q}_u^{\pi_u}(t) \mid \bm{B}_{0:T-D-1} = \bm{b}_{0:T-D-1}, \bm{C}_{0:T-D-1} = \bm{c}_{0:T-D-1} \big] . \label{Eqn:basic_emu_ideal_all_t}
\end{align}

By inserting \eqref{Eqn:basic_real_emu_Qji_t} into \eqref{Eqn:basic_emu_ideal_all_t} and taking expectation over $\bm{B}$ and $\bm{C}$, and using the fact that $\bm{B}(t)$ and $\bm{C}(t)$ are stationary over time $t$, we show that for $t \geqslant D$,
\begin{equation} \label{Eqn:basic_real_ideal_Qji}
  \mathbb{E} \big[ Q_{jk}(t) \big] = \mathbb{E} \big[ Q^{\pi_u}_{jk}(t-D) \big] + \sum_{\tau=k-D}^{t-1} A_{jk}(\tau) .
\end{equation}

By summing \eqref{Eqn:basic_real_ideal_Qji} over $t$, dividing the summation by $T$ and taking $T \to \infty$, we complete the proof.


\section{Proof of Theorem \ref{Thm:UT_Qji}} \label{App:UT_Qji}

Define the time slots when $F_{jik}(t) \geqslant Q_{jk}(t) + A_{jk}(t)$ between $D$ and $T$ as $\Gamma = \{ t_1, t_2, \cdots, t_K \}$. Then it is straightforward that
\begin{equation} \label{Eqn:UT_Qji}
    Q_{jk}(t_k+1) = 0 \leqslant Q^e_{jk}(t_k-D+1) . 
\end{equation}

We then discuss $Q_{jk}(t)$ for $D \leqslant t < t_1$. We have that in the real system,
\begin{equation} \label{Eqn:UT_Qji_2}
    Q_{jk}(t+1) = Q_{jk}(D) + \sum_{\tau = D}^{t} A_{jk}(\tau) - \sum_{\tau = D}^{t} \sum_{i \in \mathcal{M}} F_{jik}(\tau) .
\end{equation}

Meanwhile, in the emulated system, we have
\begin{equation} \label{Eqn:UT_Qji_3}
    Q^e_{jk}(t-D+1) = Q_{jk}(0) + \sum_{\tau = D}^{t} A_{jk}(\tau-D) - \sum_{\tau = D}^{t} \sum_{i \in \mathcal{M}} F_{jik}(\tau) .
\end{equation}

Comparing \eqref{Eqn:UT_Qji_2} and \eqref{Eqn:UT_Qji_3}, we have
\begin{align}
    Q_{jk}(t+1) = & Q^e_{jk}(t-D+1) + \sum_{\tau = D}^{t} \big( A_{jk}(\tau) - A_{jk}(\tau-D) \big) + Q_{jk}(D) - Q_{jk}(0) \nonumber \\
    = & Q^e_{jk}(t-D+1) + \sum_{\tau = D}^{t} \big( A_{jk}(\tau) - A_{jk}(\tau-D) \big) + \sum_{\tau = 0}^{D-1} A_{jk}(\tau) - \sum_{\tau = 0}^{D-1} \sum_{i \in \mathcal{M}} \tilde{F}_{jik}(\tau) \nonumber \\
    = & Q^e_{jk}(t-D+1) + \sum_{\tau = t-D+1}^{t} A_{jk}(\tau) - \sum_{\tau = 0}^{D-1} \sum_{i \in \mathcal{M}} \tilde{F}_{jik}(\tau) , \label{Eqn:UT_Qji_4}
\end{align}
where $\tilde{F}^D_{jik}$ denotes the number of actually served packet. 

We now discuss $Q_{jk}(t)$ for $t_k < t < t_{k+1}$. In the real system, we have
\begin{equation} \label{Eqn:UT_Qji_5}
    Q_{jk}(t+1) = Q_{jk}(t_k+1) + \sum_{\tau = t_k+1}^{t} A_{jk}(\tau) - \sum_{\tau = t_k+1}^{t} \sum_{i \in \mathcal{M}} F_{jik}(\tau) = \sum_{\tau = t_k+1}^{t} A_{jk}(\tau) - \sum_{\tau = t_k+1}^{t} \sum_{i \in \mathcal{M}} F_{jik}(\tau) .
\end{equation}

Meanwhile, in the emulated system, we have
\begin{equation} \label{Eqn:UT_Qji_6}
    Q^e_{jk}(t-D+1) = Q^e_{jk}(t_k+1-D) + \sum_{\tau = t_k+1}^{t} A_{jk}(\tau-D) - \sum_{\tau = t_k+1}^{t} \sum_{i \in \mathcal{M}} F_{jik}(\tau) .
\end{equation}

Comparing \eqref{Eqn:UT_Qji_5} and \eqref{Eqn:UT_Qji_6}, we have
\begin{align}
    Q_{jk}(t+1) = & Q^e_{jk}(t-D+1) + \sum_{\tau = t_k+1}^{t} \big( A_{jk}(\tau) - A_{jk}(\tau-D) \big) - Q^e_{jk}(t_k+1-D) \nonumber \\
    \leqslant & Q^e_{jk}(t-D+1) + \sum_{\tau = t_k+1}^{t} \big( A_{jk}(\tau) - A_{jk}(\tau-D) \big) . \label{Eqn:UT_Qji_7}
\end{align}

By summing \eqref{Eqn:UT_Qji}, \eqref{Eqn:UT_Qji_4} and \eqref{Eqn:UT_Qji_7} over $t$, dividing the summation by $T$ and taking $T \to \infty$, we have
\begin{align}
\bar{Q}_{jk} \leqslant \bar{Q}^e_{jk} + & \lim_{T \to \infty} \frac{\sum_{t=D}^{t_1} \sum_{\tau = t-D}^{t-1} A_{jk}(\tau) - (t_1-D+1) \cdot \sum_{\tau = 0}^{D-1} \sum_{i \in \mathcal{M}} \tilde{F}_{jik}(\tau)}{T} + \nonumber \\
& \lim_{T \to \infty} \frac{\sum_k \sum_{t=t_k+2}^{t_{k+1}} \sum_{\tau=t_k+1}^{t-1} \big( A_{jk}(\tau) - A_{jk}(\tau-D) \big)}{T} \nonumber \\
= \bar{Q}^e_{jk} + & \lim_{T \to \infty} \frac{\sum_{t=D}^{t_1} \min\{ D, d_{jk}(t) \} \cdot A_{jk}(t)}{T} - \lim_{T \to \infty} \frac{(t_1-D+1) \cdot \sum_{\tau = 0}^{D-1} \sum_{i \in \mathcal{M}} \tilde{F}_{jik}(\tau)}{T} + \nonumber \\
 & \lim_{T \to \infty} \frac{\sum_{t=t_1+1}^{T-1} \min\{ D, d_{jk}(t) \} \cdot A_{jk}(t)}{T} - \lim_{T \to \infty} \frac{\sum_{t=D}^{T-1} d_{jk}(t) \cdot \mathbb{1}_{e_{jk}(t) \leqslant D} \cdot A_{jk}(t-D)}{T} \nonumber \\
= \bar{Q}^e_{jk} + & \lim_{T \to \infty} \frac{\sum_{t=D}^{T-1} \min\{ D, d_{jk}(t) \} \cdot A_{jk}(t)}{T} - \lim_{T \to \infty} \frac{d_{jk}(D) \cdot \sum_{\tau = 0}^{D-1} \sum_{i \in \mathcal{M}} \tilde{F}_{jik}(\tau)}{T} - \nonumber \\
& \lim_{T \to \infty} \frac{\sum_{t=D}^{T-1} d_{jk}(t) \cdot \mathbb{1}_{e_{jk}(t) \leqslant D} \cdot A_{jk}(t-D)}{T} . \label{Eqn:UT_Qji_8}
\end{align}

By taking expectation over both sides of \eqref{Eqn:UT_Qji_8} and using the fact that $Q_{jk}$ has the same expectation between the emulated system and the ideal system as shown in the proof of Theorem \ref{Thm:basic_Qji}, we complete the proof.


\section{Proof of Theorem \ref{Thm:basic_Qib}} \label{App:basic_Qib}

We first compare $Q_{ik}(t)$ and $Q^e_{ik}(t-D)$. When $t = D$, since the servers take no action during $0 \leqslant t \leqslant D-1$, we simply have
\begin{equation} \label{Eqn:basic_real_emu_Qib_D}
    Q_{ik}(D) = Q_{ik}(0) = Q^e_{ik}(0) .
\end{equation}

Suppose when $t = w$ where $w \geqslant D$, we have
\begin{equation} \label{Eqn:basic_real_emu_Qib_k}
  Q_{ik}(w) = Q^e_{ik}(w-D) .
\end{equation}

When $t = w+1$, in the emulated system, by \eqref{Eqn:evolve_emu_Qib}, we have
\begin{equation}  \label{Eqn:basic_real_emu_Qib_kp1}
    Q^e_{ik}(w-D+1) = Q^e_{ik}(w-D) + \sum_{j \in \mathcal{N}} F_{jik}(w) - F_{isk}(w) .
\end{equation}

Meanwhile, in the real system, by \eqref{Eqn:evolve_real_Qib}, we have
\begin{align}
    Q_{ik}(w+1) = & \max \big\{ 0, Q_{ik}(w) + \sum_{j \in \mathcal{N}} F_{jik}(w) - F_{isk}(w) \big\} \nonumber \\
    = & \max \big\{ 0, Q^e_{ik}(w-D) + \sum_{j \in \mathcal{N}} F_{jik}(w) - F_{isk}(w) \big\} \nonumber \\
    = & \max \big\{ 0, Q^e_{ik}(w-D+1) \big\} \nonumber \\
    = & Q^e_{ik}(w-D+1) \label{Eqn:basic_real_emu_Qib_kp1_2} ,
\end{align}
where the second equation holds by inserting \eqref{Eqn:basic_real_emu_Qib_k} and the third equation holds by applying \eqref{Eqn:basic_real_emu_Qib_kp1}.

By combining \eqref{Eqn:basic_real_emu_Qib_D}, \eqref{Eqn:basic_real_emu_Qib_k} and \eqref{Eqn:basic_real_emu_Qib_kp1}, we have that, by induction, for $t \geqslant D$,
\begin{equation}  \label{Eqn:basic_real_emu_Qib_t}
  Q_{ik}(t) = Q^e_{ik}(t-D) .
\end{equation}

Using \eqref{Eqn:basic_emu_ideal_all_t} and taking expectation over $\bm{B}$ and $\bm{C}$, we show that for $t \geqslant D$,
\begin{equation} \label{Eqn:basic_real_ideal_Qib}
  \mathbb{E} \big[ Q_{ik}(t) \big] = \mathbb{E} \big[ Q^{\pi_u}_{ik}(t-D) \big] .
\end{equation}

By summing \eqref{Eqn:basic_real_ideal_Qib} over $t$, dividing the summation by $T$ and taking $T \to \infty$, we complete the proof.


\section{Proof of Theorem \ref{Thm:UT_Qib}} \label{App:UT_Qib}

Define the time slots when $F_{ijk}(t) \geqslant Q_{ik}(t) + \sum_{j \in \mathcal{N}} \tilde{F}_{jik}(t)$ between $D$ and $T$ as $\Gamma = \{ t_1, t_2, \cdots, t_K \}$. Then it is straightforward that
\begin{equation} \label{Eqn:UT_Qib}
    Q_{ik}(t_k+1) = 0 \leqslant Q^e_{ik}(t_k-D+1) . 
\end{equation}

We then discuss $Q_{ik}(t)$ for $D \leqslant t < t_1$. We have that in the real system,
\begin{equation} \label{Eqn:UT_Qib_2}
    Q_{ik}(t+1) = Q_{ik}(D) + \sum_{\tau = D}^{t} \sum_{j \in \mathcal{N}} \tilde{F}_{jik}(\tau) - \sum_{\tau = D}^{t} F_{isk}(\tau) .
\end{equation}

Meanwhile, in the emulated system, we have
\begin{equation} \label{Eqn:UT_Qib_3}
    Q^e_{ik}(t-D+1) = Q_{ik}(0) + \sum_{\tau = D}^{t} \sum_{j \in \mathcal{N}} \tilde{F}_{jik}(\tau) - \sum_{\tau = D}^{t} F_{isk}(\tau) .
\end{equation}

Comparing \eqref{Eqn:UT_Qib_2} and \eqref{Eqn:UT_Qib_3}, we have
\begin{align} 
    Q_{ik}(t+1) = & Q^e_{ik}(t-D+1) + Q_{ik}(D) - Q_{ik}(0) \nonumber \\
    = & Q^e_{ik}(t-D+1) + \sum_{\tau = 0}^{D-1} \sum_{j \in \mathcal{N}} \tilde{F}_{jik}(\tau) - \sum_{\tau = 0}^{D-1} \tilde{F}_{isk}(\tau) . \label{Eqn:UT_Qib_4}
\end{align}

We now discuss $Q_{ik}(t)$ for $t_k < t < t_{k+1}$. In the real system, we have
\begin{align}
    Q_{ik}(t+1) = & \ Q_{ik}(t_k+1) + \sum_{\tau = t_k+1}^{t} \sum_{j \in \mathcal{N}} \tilde{F}_{jik}(\tau) - \sum_{\tau = t_k+1}^{t} F_{isk}(\tau) \nonumber \\
    = & \sum_{\tau = t_k+1}^{t} \sum_{j \in \mathcal{N}} \tilde{F}_{jik}(\tau) - \sum_{\tau = t_k+1}^{t} F_{isk}(\tau) . \label{Eqn:UT_Qib_5}
\end{align}

Meanwhile, in the emulated system, we have
\begin{equation} \label{Eqn:UT_Qib_6}
    Q^e_{ik}(t-D+1) = Q^e_{ik}(t_k+1-D) + \sum_{\tau = t_k+1}^{t} \sum_{j \in \mathcal{N}} \tilde{F}_{jik}(\tau) - \sum_{\tau = t_k+1}^{t} F_{isk}(\tau) . 
\end{equation}

Comparing \eqref{Eqn:UT_Qib_5} and \eqref{Eqn:UT_Qib_6}, we have
\begin{equation} \label{Eqn:UT_Qib_7}
  Q_{ik}(t+1) = Q^e_{ik}(t-D+1) - Q^e_{ik}(t_k+1-D) \leqslant Q^e_{ik}(t-D+1) .
\end{equation}

By summing \eqref{Eqn:UT_Qib}, \eqref{Eqn:UT_Qib_4} and \eqref{Eqn:UT_Qib_7} over $t$, dividing the summation by $T$ and taking $T \to \infty$, we have
\begin{align} 
  \bar{Q}_{ik} \leqslant & \bar{Q}^e_{ik} + \lim_{T \to \infty} \frac{(t_1-D+1) \cdot \big( \sum_{\tau = 0}^{D-1} \sum_{j \in \mathcal{N}} \tilde{F}_{jik}(\tau) - \sum_{\tau = 0}^{D-1} \tilde{F}_{isk}(\tau) \big)}{T} \nonumber \\
  = & \bar{Q}^e_{ik} + \sum_{t = 0}^{D-1} \Big( \sum_{j \in \mathcal{N}} \tilde{F}_{jik}(t) - \tilde{F}_{isk}(t) \Big) \cdot \lim_{T \to \infty} \frac{d_{ik}(D)}{T} \label{Eqn:UT_Qib_8}
\end{align}

By taking expectation over both sides of \eqref{Eqn:UT_Qib_8} and following the similar analysis as the proof of Theorem \ref{Thm:basic_Qib}, we complete the proof.


\section{Proof of Theorem \ref{Thm:basic_Qju}} \label{App:basic_Qju}

We first compare $Q_{jk}(t)$ and $Q^e_{jk}(t-D)$. Equations from \eqref{Eqn:basic_real_emu_Qju_D} to \eqref{Eqn:basic_real_emu_Qju_kp1_2} are all under the condition that $\bm{B}_{D:T-1} = \bm{b}_{0:T-D-1}$ for the real system and $\bm{B}_{0:T-D-1} = \bm{b}_{0:T-D-1}$ for the emulated system.

When $t = D$, since the dispatchers take no action during $0 \leqslant t \leqslant D-1$, and thus no packets arrive at INP $j$ during the period. Meanwhile, INP $j$ can serve buffered packets, thus we have
\begin{equation} \label{Eqn:basic_real_emu_Qju_D}
    Q_{jk}(D) \leqslant Q_{jk}(0) = Q^e_{jk}(0).
\end{equation}

Suppose when $t = w$ where $w \geqslant D$, we have 
\begin{equation} \label{Eqn:basic_real_emu_Qju_k}
    Q_{jk}(w) \leqslant Q^e_{jk}(w-D) .
\end{equation}

When $t = w+1$, in the emulated system, by \eqref{Eqn:evolve_emu_Qju}, we have
\begin{align} 
   Q^e_{jk}(w-D+1) = & \max \Big\{ 0, \ Q^e_{jk}(w-D) + \sum_{i \in \mathcal{M}} F_{ijk}(w) - B_{jsk}(w-D) \Big\} \nonumber \\
   = & \max \Big\{ 0, \ Q^e_{jk}(w-D) + \sum_{i \in \mathcal{M}} F_{ijk}(w) - b_{j, k-D} \Big\} . \label{Eqn:basic_real_emu_Qju_kp1}
\end{align}

Meanwhile, in the real system, by \eqref{Eqn:evolve_real_Qju}, we have
\begin{align} 
    Q_{jk}(w+1) = & \max \Big\{ 0, \ Q_{jk}(w-D) + \sum_{i \in \mathcal{M}} F_{ijk}(w-D) - B_{jsk}(w) \Big\} \nonumber \\
    = & \max \Big\{ 0, \ Q_{jk}(w-D) + \sum_{i \in \mathcal{M}} F_{ijk}(w-D) - b_{j, k-D} \Big\} \nonumber \\
    \leqslant & \max \Big\{ 0, \ Q^e_{jk}(w-D) + \sum_{i \in \mathcal{M}} F_{ijk}(w-D) - b_{j, k-D} \Big\} \nonumber \\
    = & Q^e_{jk}(w-D+1) , \label{Eqn:basic_real_emu_Qju_kp1_2}
\end{align}
where the inequality holds by applying \eqref{Eqn:basic_real_emu_Qju_k} and the last equation holds by applying \eqref{Eqn:basic_real_emu_Qju_kp1}.

By combining \eqref{Eqn:basic_real_emu_Qju_D}, \eqref{Eqn:basic_real_emu_Qju_k} and \eqref{Eqn:basic_real_emu_Qju_kp1_2}, we have that, by induction, for $t \geqslant D$,
\begin{equation} \label{Eqn:basic_real_emu_Qju_t}
    \mathbb{E} \big[ Q_{jk}(t) \mid \bm{B}_{D:T-1} = \bm{b}_{0:T-D-1} \big] \leqslant \mathbb{E} \big[ Q^e_{jk}(t-D) \mid \bm{B}_{0:T-D-1} = \bm{b}_{0:T-D-1} \big] .    
\end{equation}

Using \eqref{Eqn:basic_emu_ideal_all_t} and \eqref{Eqn:basic_real_emu_Qju_t}, we have that
\begin{align}
    & \mathbb{E} \big[ \bm{Q}(t) \mid \bm{B}_{D:T-1} = \bm{b}_{0:T-D-1}, \bm{C}_{D:T-1} = \bm{c}_{0:T-D-1} \big] \nonumber \\
    \leqslant & \mathbb{E} \big[ \bm{Q}^e(t) \mid \bm{B}_{0:T-D-1} = \bm{b}_{0:T-D-1}, \bm{C}_{D:T-1} = \bm{c}_{0:T-D-1} \big] \nonumber \\
    = & \mathbb{E} \big[ \bm{Q}^{\pi_d}(t) \mid \bm{B}_{0:T-D-1} = \bm{b}_{0:T-D-1}, \bm{C}_{0:T-D-1} = \bm{c}_{0:T-D-1} \big] . \label{Eqn:basic_real_ideal_Qju_t}
\end{align}

By taking expectation over $\bm{B}$ and $\bm{C}$ in \eqref{Eqn:basic_real_ideal_Qju_t}, and using the fact that $\bm{B}(t)$ and $\bm{C}(t)$ are stationary over time $t$, we show that for $t \geqslant D$,
\begin{equation} \label{Eqn:basic_real_ideal_Qju}
  \mathbb{E} \big[ Q_{jk}(t) \big] \leqslant \mathbb{E} \big[ Q^{\pi_d}_{jk}(t-D) \big] .
\end{equation}

By summing \eqref{Eqn:basic_real_ideal_Qju} over $t$, dividing the summation by $T$ and taking $T \to \infty$, we complete the proof.


\section{Proof of Theorem \ref{Thm:UT_Qju}} \label{App:UT_Qju}

Equations from \eqref{Eqn:UT_Qju_2} to \eqref{Eqn:UT_Qju_8} are all under the condition that $\bm{B}_{D:T-1} = \bm{b}_{0:T-D-1}$ for the real system and $\bm{B}_{0:T-D-1} = \bm{b}_{0:T-D-1}$ for the emulated system.

Define the time slots when $b_{j, t-D} \geqslant Q_{jk}(t) + \sum_{i \in \mathcal{M}} \tilde{F}_{ijk}(t)$ between $D$ and $T$ as $\Gamma = \{ t_1, t_2, \cdots, t_K \}$. Then it is straightforward that
\begin{equation} \label{Eqn:UT_Qju}
    Q_{jk}(t_k+1) = 0 \leqslant Q^e_{jk}(t_k-D+1) . 
\end{equation}

We then discuss $Q_{jk}(t)$ for $D \leqslant t < t_1$. We have that in the real system,
\begin{equation} \label{Eqn:UT_Qju_2}
    Q_{jk}(t+1) = Q_{jk}(D) + \sum_{\tau = D}^{t} \sum_{i \in \mathcal{M}} \tilde{F}_{ijk}(\tau) - \sum_{\tau = D}^{t} b_{j, \tau-D} .
\end{equation}

Meanwhile, in the emulated system, we have
\begin{equation} \label{Eqn:UT_Qju_3}
    Q^e_{jk}(t-D+1) = Q_{jk}(0) + \sum_{\tau = D}^{t} \sum_{i \in \mathcal{M}} \tilde{F}_{ijk}(\tau) - \sum_{\tau = D}^{t} b_{j, \tau-D} .
\end{equation}

Comparing \eqref{Eqn:UT_Qju_2} and \eqref{Eqn:UT_Qju_3}, we have
\begin{align} 
    Q_{jk}(t+1) = & Q^e_{jk}(t-D+1) + Q_{jk}(D) - Q_{jk}(0) \nonumber \\
    = & Q^e_{ik}(t-D+1) + \sum_{\tau = 0}^{D-1} \sum_{i \in \mathcal{M}} \tilde{F}_{ijk}(\tau) - \sum_{\tau = 0}^{D-1} \tilde{B}_{jsk}(\tau) . \label{Eqn:UT_Qju_4}
\end{align}

We now discuss $Q_{jk}(t)$ for $t_k < t < t_{k+1}$. In the real system, we have
\begin{equation} \label{Eqn:UT_Qju_5}
    Q_{jk}(t+1) = Q_{jk}(t_k+1) + \sum_{\tau = t_k+1}^{t} \sum_{i \in \mathcal{M}} \tilde{F}_{ijk}(\tau) - \sum_{\tau = t_k+1}^{t} b_{j, \tau-D} = \sum_{\tau = t_k+1}^{t} \sum_{i \in \mathcal{M}} \tilde{F}_{ijk}(\tau) - \sum_{\tau = t_k+1}^{t} b_{j, \tau-D} .
\end{equation}

Meanwhile, in the emulated system, we have
\begin{equation} \label{Eqn:UT_Qju_6}
    Q^e_{jk}(t-D+1) = Q^e_{jk}(t_k+1-D) + \sum_{\tau = t_k+1}^{t} \sum_{i \in \mathcal{M}} \tilde{F}_{ijk}(\tau) - \sum_{\tau = t_k+1}^{t} b_{j, \tau-D} .
\end{equation}

Comparing \eqref{Eqn:UT_Qju_5} and \eqref{Eqn:UT_Qju_6}, we have
\begin{equation} \label{Eqn:UT_Qju_7}
  Q_{jk}(t+1) = Q^e_{jk}(t-D+1) - Q^e_{jk}(t_k+1-D) \leqslant Q^e_{jk}(t-D+1) .
\end{equation}

By summing \eqref{Eqn:UT_Qju}, \eqref{Eqn:UT_Qju_4} and \eqref{Eqn:UT_Qju_7} over $t$, dividing the summation by $T$ and taking $T \to \infty$, we have
\begin{align} 
  \bar{Q}_{ik} \leqslant & \bar{Q}^e_{ik} + \lim_{T \to \infty} \frac{(t_1-D+1) \cdot \big( \sum_{\tau = 0}^{D-1} \sum_{i \in \mathcal{M}} \tilde{F}_{ijk}(\tau) - \sum_{\tau = 0}^{D-1} \tilde{B}_{jsk}(\tau) \big)}{T} \nonumber \\
  = & \bar{Q}^e_{ik} + \sum_{t = 0}^{D-1} \Big( \sum_{i \in \mathcal{M}} \tilde{F}_{ijk}(t) - \tilde{B}_{jsk}(t) \Big) \cdot \lim_{T \to \infty} \frac{d_{jk}(D)}{T} \label{Eqn:UT_Qju_8}
\end{align}

By taking expectation over both sides of \eqref{Eqn:UT_Qju_8} and following the similar analysis as the proof of Theorem \ref{Thm:basic_Qju}, we complete the proof.


\end{document}